\documentclass[article]{aa} 

\usepackage{graphicx}
\usepackage{caption}
\usepackage{subcaption}
\usepackage{epstopdf}
\usepackage{amsmath}
\usepackage[english]{babel}
\usepackage[utf8]{inputenc}
\usepackage{amssymb}
\usepackage{wrapfig}
\usepackage{pdflscape}
\usepackage{lscape}
\usepackage{rotating}
\usepackage{longtable}
\usepackage{appendix}
\usepackage{textcomp}
\usepackage{multirow}
\bibpunct{(}{)}{;}{a}{}{,}
\usepackage{array}
\usepackage{ragged2e}
\usepackage{adjustbox}
\usepackage{txfonts}
\usepackage{xcolor}
\usepackage{float}
\usepackage[bookmarksnumbered=true]{hyperref} 
\hypersetup{
    colorlinks = true,
    linkcolor = blue,
    anchorcolor = blue,
    citecolor = blue,
    filecolor = blue,
    urlcolor = blue
    }
\usepackage{titlesec}
\usepackage{comment}
\usepackage{lscape}

\begin{document}

   \title{A comprehensive radio study of narrow-line Seyfert 1 galaxies}
 \titlerunning{NLS1s in radio}

   \author{I. Varglund\inst{1,2}\thanks{irene.varglund@aalto.fi}
          \and
          E. Järvelä\inst{3}\thanks{Dodge Family Prize Fellow at The University of Oklahoma}
          \and
          M. J. Hardcastle\inst{4}
          \and
          S. Varglund\inst{5}
          \and
          A. Lähteenmäki\inst{1,2}
          }
            \institute{Aalto University Metsähovi Radio Observatory, Metsähovintie 114, FI-02540 Kylmälä, Finland
            \and
             Aalto University Department of Electronics and Nanoengineering, P.O. Box 15500, FI-00076 AALTO, Finland
             \and
             Homer L. Dodge Department of Physics and Astronomy, The University of Oklahoma, 440 W. Brooks St., Norman, OK 73019, USA
             \and
             Centre for Astrophysics Research, University of Hertfordshire, College Lane, Hatfield AL10 9AB, UK
             \and
             Private researcher, Raisio, Finland}

   \date{Received }

  \abstract
{Narrow-line Seyfert 1 (NLS1) galaxies are a type of active galactic nuclei (AGN) that were originally classified as sources with little to no radio emission. Although the class is rather unified from an optical perspective, their radio characteristics are diverse. One of the biggest curiosities found in these sources is their ability to form and maintain powerful relativistic jets. We studied the radio properties of a sample of 3998 NLS1 galaxies which is the largest clean sample available, thus allowing us to study the population-wide characteristics. We used both historical and ongoing surveys: the LOw-Frequency ARray (LOFAR) Two-metre Sky Survey (LoTSS; 144~MHz), the Faint Images of the Radio Sky at Twenty-centimeters (FIRST; 1.4~GHz), the National Radio Astronomy Observatory (NRAO) Very Large Array (VLA) Sky Survey (NVSS; 1.4~GHz), and the VLA Sky Survey (VLASS; 3~GHz). We were able to obtain a radio detection for $\sim40\%$ of our sources, with the most detections by LoTSS. The majority of the detected NLS1 galaxies are faint ($\sim1-2$ mJy) and non-variable, suggesting considerable contributions from star formation activities, especially at 144~MHz. However, we identified samples of extreme sources, for example, in fractional variability and radio luminosity, indicating significant AGN activity. Our results highlight the heterogeneity of the NLS1 galaxy population in radio and lays the foundation for targeted future studies.

}

  \keywords{galaxies: active -- galaxies: Seyfert -- galaxies: jets -- galaxies: star formation -- galaxies: statistics -- radio continuum: galaxies}

   \maketitle
%

\section{Introduction}
\label{sec:intro}

Narrow-line Seyfert 1 (NLS1) galaxies \citep{1985osterbrock1} are a class of active galactic nuclei (AGN) with perplexingly peculiar properties. The classical way of identifying NLS1 galaxies is through the full width at half maximum (FWHM) of their H$\beta$ line, which is by definition less than 2000~km s$^{-1}$ \citep[]{1989goodrich1}. Further identifying properties from an optical point of view include weak [O~\textsc{III}] emission with respect to H$\beta$ ($\rm S([O~\textsc{III}]\lambda5007) / S(H\beta) < 3$ \citealp{1985osterbrock1}) and strong Fe~\textsc{II} multiplets found in a considerable fraction of NLS1 galaxies \citep{1992boroson}. The former is also a classification criterion, ensuring that the source is a Type 1 AGN. A final optical trait is found in the blueshifted line profiles identified in some NLS1 galaxies, a feature commonly only found in high-ionization lines and indicative of outflows \citep[e.g.][]{2002zamanov,2005boroson,2008komossa,2021berton2}.  

From the physical point of view, the low-mass black hole plays a key role in understanding the curious nature of this AGN class. The mass of the central supermassive black hole of NLS1 galaxies is on average $M_{\text{BH}} \sim 10^{6} M_{\sun}$ - $10^{8} M_{\sun}$; \citealp[e.g.][]{2011peterson, 2015jarvela, 2016cracco1, 2018chen}, as confirmed by reverberation mapping campaigns \citep[e.g.][]{2016wang, 2018du, 2019du1}. The class-defining narrow H$\beta_{broad}$ line is commonly attributed to the broad-line region (BLR) clouds having low orbital velocities due to the low mass of the central black hole \citep{2011peterson}. Beyond the low black hole mass, another key physical property exhibited by NLS1 galaxies is the high Eddington ratio. On average, the Eddington ratio is between 0.1 and 1, however, there are some cases exceeding unity \citep{1992boroson, 2018marziani, 2022tortosa}. Furthermore, NLS1 nuclei are preferentially hosted in disk-like galaxies \citep{2018jarvela, 2022varglund,2023varglund} residing in low-density Mpc-scale environments \citep{2017jarvela}. With all these properties in mind, the common understanding is that NLS1 galaxies represent an early stage in AGN evolution \citep{2000mathur}.

The rate of NLS1 galaxies detected at 1.4~GHz using FIRST or NVSS has varied significantly, from $\sim7\%$ to $\sim30\%$ \citep{2006komossa1,2006zhou}. Even lower detection rates have been noted, $\sim5\%$ ($\sim11000$ sources) \citep{2017rakshit} and $\sim3\%$ ($\sim22000$ sources) \citep{2024paliya}. The larger occurrence of higher redshift sources in \citet{2024paliya} in comparison to, for example, \citet{2006komossa1} has been provided as an explanation for the lower detection rate. However, the contamination rate of \citet{2017rakshit} is high, as evidenced by \citet{2020berton2} discarding more than half of the sources in \citet{2017rakshit} as unreliable or misclassifications. This holds true for \citet{2024paliya} as well, as $\sim87\%$ of the sources in \citet{2017rakshit} are also in \citet{2024paliya}.

In most of the sources in \citet{2006komossa1} the radio emission dominates over the optical emission. This class of AGN is capable of harboring powerful relativistic jets, similar to those found in blazars \citep[e.g.][]{2006komossa1, 2008yuan, 2011foschini, 2015foschini1, 2017lahteenmaki1, 2018lahteenmaki1}. The existence of jets in these predominantly disk-like sources with low-to-intermediate mass supermassive black holes directly contradicts the jet paradigm, according to which only sources with $M_{BH} > 10^8 M_\sun$ can launch powerful relativistic jets \citep{2000laor1}. The jetted NLS1 galaxies show strong variability, due to the changes in the jet and enhanced by Doppler boosting, especially in radio and gamma-rays \citep{2017huang}. Otherwise, the jetted NLS1 population does not significantly differ from the non-jetted one in, for example, host galaxy morphologies \citep{2018jarvela,2022varglund,2023varglund}.

From an optical perspective the NLS1 galaxy population is very unified, in other aspects, such as from the radio perspective, the population is mixed at best \citep[e.g.,][]{2015jarvela,2018marziani,2022jarvela}. Due to misclassifications, a result of, for example, using fully automated spectral modeling algorithms and identifying processes combined with poor-quality data, a large fraction of sources claimed to be NLS1 galaxies in the literature are not \citep{2015sulentic1,2017rakshit,2020berton2}. Intermediate-type AGN and broad-line Seyfert 1 (BLS1) galaxies are the most common culprits of contamination in NLS1 galaxy samples. BLS1 galaxies are different from NLS1 galaxies in several aspects, such as in their evolutionary stage, Eddington ratios, and black hole masses, as well as in the variability and multiwavelength properties \citep{2004klimek,2011orban,2017jarvela,2020gliozzi,2023wang}. Due to the contamination in previous samples, it has been impossible to study NLS1 galaxies statistically as a population.

With the earlier samples of NLS1 galaxies being contaminated, there is no clear understanding of the population-wide radio properties. Furthermore, due to the detection thresholds of former surveys being so high, a vast amount of NLS1 galaxies have simply remained undetected. Newer surveys offer better sensitivity and thus a lower detection threshold which will give a better idea of the true detection rate of these sources. Due to the nature and properties of NLS1 galaxies, it is difficult to find catalogs with large samples of these sources. The NLS1 galaxy population-wide properties are mostly unknown with only a few studies reporting radio properties of large NLS1 galaxy samples \citep{2006komossa1, 2022jarvela}, nonetheless these sources do present the most diversity in radio.  Due to the contamination of earlier samples, it has not been possible to draw meaningful, statistical conclusions on the radio properties of the class. With the sample used in this study, further discussed below, and the results, as well as with other new radio surveys, we can finally obtain population-wide results of this peculiar class of AGN. By using radio, we can look at these sources using historical catalogs as well as more modern ones. This can provide clues on the AGN behavior, such as variability. Another key benefit of using radio is that it can give us information on a variety of different properties of NLS1 galaxies, including, for example, the radio emission mechanism(s) and the possible presence of relativistic jets. 

The sample used in the present paper is the cleanest large sample currently available. To produce this sample, we first removed all the sources with signal-to-noise (S/N) < 5 in the $\lambda$5100~Å continuum, since it has been shown that at low S/N it is impossible to distinguish NLS1 galaxies from sources that have narrow H$\beta$ line due to other reasons, such as obscuration \citep{2020jarvela}. Then we fit the H$\beta$ line with both a Lorentzian and Gaussian profile and performed a manual inspection of the results and removed additional sources that could visually be identified as intermediate sources or that clearly did not have high-quality enough data to perform reliable modeling. The main goal of this analysis was to build a sample of sources that we can trust to be NLS1 galaxies. A detailed explanation of how the sample was produced with all the used selection criteria can be found in \citet{2020berton2}.

We have carried out an extensive study of this sample in radio. The surveys we have made use of in this paper are Faint Images of the Radio Sky at Twenty-centimeters (FIRST) \citep{2015helfand}, the National Radio Astronomy Observatory (NRAO) Very Large Array (VLA) Sky Survey (NVSS) \citep{1998condon}, the VLA Sky Survey (VLASS) \citep{2021gordon}, and the LOw Frequency ARray (LOFAR) Two-metre Sky Survey (LoTSS) \citep{2022shimwell}. The central frequencies of the studied surveys are 144~MHz (LoTSS), 1.4~GHz (FIRST and NVSS), and 3~GHz (VLASS). A description of the sample and how we determined what surveys to use, as well as a short description of each survey can be found in Sect.~\ref{sec:sample}. The data analysis process and each of the studied key parameters are explained in Sect.~\ref{sec:dataanalysis}. The results can be found in Sect.~\ref{sec:dataanalysis} and a discussion of the results in Sect.~\ref{sec:discussion}. Conclusions are described in Sect.~\ref{sec:conclusions}. In this paper we assume standard $\lambda$CDM cosmology with parameters H$_{0}$ = 73 km s$^{-1}$ Mpc$^{-1}$, $\Omega_{\text{matter}}$ = 0.27, and $\Omega_{\text{vacuum}}$ = 0.73 \citep{2007spergel1}.

\section{Sample and survey selection}
\label{sec:sample}

Our sample is selected from \citet{2017rakshit}, using the selection criteria from \citet{2020berton2}, resulting in 3998 NLS1 galaxies located in the northern hemisphere. This sample is the largest clean sample of NLS1 galaxies available and is a pathway for a deeper understanding of the class. As mentioned in Sect.\ref{sec:intro}, misclassifications plague NLS1 galaxy research leading to skewed, or even wrong, conclusions. One of the aims of this study is to see how, for example, the detection rate of this sample compares to former studies that used contaminated samples. Our set of NLS1 galaxies has been thoroughly examined to avoid contamination as much as possible. This set is not perfect as there most likely are some BLS1 galaxies and some intermediate Seyfert galaxies in the sample. Obtaining a sample set with no contaminators is not possible with the current data quality as, for example, discerning between a true NLS1 and an intermediate can be quite cumbersome, even with high-quality data \citep{2020berton,2021jarvela}. Furthermore, the existence of changing-look NLS1 galaxies does not ease the process \citep{2024xu}. Nonetheless, the contamination fraction of our sample is minimal compared to the other publicly available NLS1 samples, as, for example, we were able to eliminate $\sim64\%$ of the \citet{2017rakshit} sample as most likely not NLS1 galaxies or sources that cannot reliably be identified. To obtain a purely clean NLS1 galaxy sample significantly better quality data is needed as well as a more scrutinized data processing methods.

\begin{table}
\caption{Basic properties of the sample}
\centering
\begin{tabular}{ l l l l l}
\hline\hline
 & Min    & Max    & Mean & Median \\ \hline
$z_{All}$                           & 0.01 & 0.80 & 0.34  & 0.30 \\
$z_{Det}$                           & 0.02 & 0.80 & 0.32  & 0.28 \\
log($M_{BH, \texttt{all}}$)         & 5.07 & 7.99 & 6.98  & 7.02 \\
log($M_{BH, \texttt{det}}$)         & 5.31 & 7.90 & 6.98  & 7.02 \\
FWHM(H$\beta$) [km s$^{-1}$]          & 312    & 2200   & 1710    & 1840   \\
L5100 [$10^{-7}$ W]                   & 41.76  & 45.66  & 43.73   & 43.75  \\
F(H$\beta$) [$10^{-20}$ W m$^{-2}$]   & 6      & 11606  & 476     & 268   \\
F([O III]) [$10^{-20}$ W m$^{-2}$]    & 1      & 9231   & 201     & 93    \\
R4570                       & 0.01   & 4.51   & 0.38    & 0.24  \\
Eddington ratio             & 0.03   & 11.05  & 0.56    & 0.38 
\\\hline
\end{tabular}
\label{tab:basic}
\end{table}

We have collected both archival radio detections as well as more recent radio detections of these sources, using 5-$\sigma$ as the detection threshold in all cases. Basic information regarding our sample can be found in Table~\ref{tab:basic}. We note that though simultaneous data would be ideal, obtaining it for a sample of 3998 sources at several radio frequencies would be extremely hard, and thus, the data used in this study are not simultaneous. The surveys in this study were chosen as they cover much of the northern hemisphere, where all of our sources are located, and are the most extensive radio surveys of the northern hemisphere at the time of writing. Further explanations of the catalog choices can be found in the relevant sections of the paper.

The frequencies and frequency ranges in this study are 120-168~MHz (LoTSS), 1.4~GHz (FIRST and NVSS), and 2-4~GHz (VLASS). For this study, we use data release (DR) 14 for FIRST and DR 3 for LoTSS. The bandwidth of FIRST is 42~MHz and the bandwidth of NVSS is 84~MHz. Furthermore, we include both VLASS epoch 1 and VLASS epoch 2. By studying multiple epochs within the same survey we can investigate possible variability of our sources with the same instrument setup, thus avoiding potential problems such as changes in resolution and/or data analysis methods. We are also interested in variability between the different catalogs, but note that in these cases factors like length of time between the two observations, resolution, spectrum of the source, extendedness of the emission of the source, and frequency differences must be taken into account. The time frame for this data set spans roughly three decades. 

As these sources are NLS1 galaxies and thus present interesting optical spectral properties, thus, we have also obtained spectral measurements from \citet{2017rakshit} for these sources. As this is also the paper from which our sample originates, we have the optical spectral measurements for almost all sources. We do note that due to issues in \citet{2017rakshit}, some of the spectral results might be of inadequate quality. However, thanks to the used S/N ratio criteria, these 3998 sources have adequate-quality optical spectra, and probably also better-than-average measurements from the spectra. These results should be of good enough quality to enable general studies on the possible relations. We have calculated the black hole masses of our sources with the spectral information obtained from \citet{2017rakshit}. The black hole mass is calculated using
\begin{equation}
    M_{BH} = f\frac{v^2R_{BLR}}{G},
\end{equation}
where $v$ is the rotational velocity of the gas that surrounds the black hole. The radius of the BLR has been calculated by using the relation from \citet{2013bentz} and $\lambda5100$~Å continuum luminosity. We have assumed a spherical distribution of clouds so that $f$ is $3/4$. We have assumed virialized gas.

\subsection{FIRST}

The story of large-scale 1.4~GHz surveys begins in 1990 with FIRST with the idea of creating a similar scale survey as the Palomar Observatory Sky Survey carried out in the 1950s, but this time in the radio \citep{1994becker}. The FIRST survey covers an area of over 10 000 deg$^2$ with the majority of this region also coinciding with the coverage of the Sloan Digital Sky Survey (SDSS). The sky coverage is optimal for our sample, as all of our sources are selected from SDSS. The survey uses two narrow-band observing windows at 1365~MHz and 1435~MHz or at 1335~MHz and 1730~MHz depending on the data release. The detection threshold is 1~mJy, the resolution is 5", and the average root-mean-square (rms) is 0.15~mJy. The noise is quite uniform throughout the survey, varying only $\sim$15\% between best and worst regions, when no bright sources ($>$ 100~mJy) are nearby. The data release used in this paper is the latest data release, which means that the two frequencies used by FIRST are 1335~MHz and 1730~MHz, the bandpass is 2$\times$64 2~MHz channels, and the integration time is 60~seconds. With the 1~mJy detection threshold, FIRST provides a great tool for studying NLS1 galaxies as the majority of these sources tend to, based on earlier samples, present flux densities of a few mJy \citep[e.g.,][]{2015jarvela,2020fan}.

\subsection{NVSS}

The NVSS survey \citep{1998condon} was proposed around the same time as FIRST, with the observations running from 1993 until 1996, covering the sky north of declination $-$40 degrees. As with FIRST, the NVSS observations are carried out in two windows centered at 1365~MHz and 1435~MHz. The FWHM angular resolution is $45"$, and the detection threshold of NVSS is $\sim$2.5~mJy. Due to the significantly higher detection threshold compared to FIRST, a lot of NLS1 galaxies end up not being detected with NVSS. Due to FIRST having a better resolution and a lower detection threshold, FIRST is in most cases our primary archival 1.4~GHz survey. However, as a consequence of NVSS's much wider sky coverage, some sources have been detected with NVSS and not with FIRST, hence why the survey is included in this work.

\subsection{LoTSS}

The lowest frequency range survey considered in this study is provided by LoTSS \citep{2017shimwell}, with a center frequency of 144~MHz.  The LoTSS survey is ongoing, and its next data release, DR 3, will cover all of the northern, high-Galactic-latitude sky, with a total sky area around 19,000 deg$^2$ of contiguous coverage including some regions in the Galactic plane. The DR 3 processing strategy is broadly the same as that described by \cite{2022shimwell} and will be described further by Shimwell et al. (in prep.). We based our analysis on the first internal data release of DR3, which meant that 3,725 of our sample sources were visible to LoTSS. As part of the internal data release PyBDSF is used to extract a source catalog for the whole DR 3 area using a 4.5 $\sigma$ detection threshold. However, for consistency in this paper we only consider sources with $S_{\texttt{int}}/\texttt{rms}>5$. We then crossmatched with the positions of our NLS1 targets with a search radius of $5''$. With these criteria, we obtained 1,660 LOFAR detections. When taking into account the sky coverage of LoTSS DR 3, $\sim 44.6$\% of our sources were detected by this survey.

Each catalog had sources that were only detected in that catalog. The numbers of sources detected in each survey can be found in Table~\ref{tab:detection}.

\subsection{VLASS}

The final and highest frequency radio survey included in this work is the ongoing VLASS survey. The survey began in 2017 and will consist of three epochs, two of which are available at the time of writing this paper. The first epoch was carried out between 2017 and 2019 and the second epoch from 2020 to 2022. The third epoch is currently ongoing. The survey covers everything above a declination of $-$40 degrees and is run by the NRAO.

We are using data from the Quick Look catalog which uses single epoch imaging. The synthesized beam size is $\sim$2.5" and the median rms is 120 $\mu$Jy beam$^{-1}$. The project aims to obtain an rms of 70 $\mu$Jy beam$^{-1}$ by co-adding images from all three epochs when available. The positional accuracy of the Quick Look images is $\sim$1" when the declination is southward of $-$20 degrees. With a larger than $-$ 20-degree declination, the positional accuracy is $\sim$0.5". The largest detectable angular scale of VLASS is not a limiting factor for our sources as the size of our sources is small enough that even at low redshifts the entire galaxy remains below the threshold, $\geq 30"$, at which diffuse emission is resolved out \citep{2020lacy}. For the lowest redshift source, $z = 0.0132$, this translates to a physical scale of 8.1~kpc. We do not expect to have significant amount of radio emission at scales larger than this.

At the time of writing, we have version 1.2 available for epoch 1. With version 1.2, the peak flux density values are systematically $\sim$8\% too low \footnote{For more information see the 2022 catalog documentation in the User Storage at https://cirada.ca/vlasscatalogueql0}. The total flux densities have a similar offset of $\sim$3\% in version 1.2. The astrometric error in epoch 1 is up to $\sim$1".

The epoch 2 data is at the time of the analysis incomplete, but the parameters we are interested in are already included in the catalog. The flux underestimation in epoch 2 is $\sim$8\% \footnote{For more information see https://science.nrao.edu/vlass/data-access/vlass-epoch-1-quick-look-users-guide}. We do not take the flux underestimations into account in this study. In cases in which the peak flux of a source is less than 3~mJy, there may be no counterpart in the other catalog. This happens due to a poor S/N ratio. In epoch 2, $\sim$7\% of single epoch sources with flux densities below 3~mJy do not have a counterpart in epoch 1 images. The same happens the opposite way in $\sim$4\% of the cases. Due to several NLS1 galaxies having low flux densities at this frequency, we may run into some missing counterparts. However, in some cases this can also be a result of intrinsic variability.

\begin{table*}
\caption{Detections and detection rate for each catalog}
\centering
\begin{tabular}{ l l l l l l}
\hline\hline
Catalog & Detections & Detection rate$^a$ & Median rms$^b$ & Angular resolution \\ 
        &          & \% &         & mJy beam$^{-1}$ &   \\ \hline
FIRST               & 302     & 7.6 & 0.75  & $\sim5"$ \\ 
NVSS                & 88      & 2.2 & 2.25  & $\sim45"$ \\
LoTSS DR 3          & 1660    & 44.6 & $0.09^c$ & $\sim6"^d$  \\ 
VLASS Epoch 1       & 211     & 5.3 & 0.60  & $\sim15"$ \\ 
VLASS Epoch 2       & 206     & 5.2 & 0.60  & $\sim15"$ \\ 
\\\hline
\end{tabular}
\tablefoot{ a) The detection rate is given based on the whole sample when taking into account survey coverage. \\ b) The sensitivity is given for 5$\sigma$. \\ c) This is the sensitivity of the data for our sample. Overall, the sensitivity of DR 3 ranges between 0.07 and 0.3 mJy beam$^{-1}$ depending on the observational parameters.  \\ d) The given sensitivity is for declinations $>\sim10\deg$, for lower declinations the sensitivity is $\sim9"$.}
\label{tab:detection}
\end{table*}

\section{Radio data analysis and results}
\label{sec:dataanalysis}

To study the possible matches from various catalogs we used a search radius of 5". For this paper, we are only including sources that have been detected at least in one of the included surveys. With the used search radius, we obtained at least one radio detection for $\sim40\%$ of the sample (see Table~\ref{tab:detection}). In the case of having a detection with both FIRST and NVSS, we favored the FIRST result due to its better resolution. This section will only cover the data analysis and results. The interpretation and deeper analysis of what these results mean can be found in Sect.~\ref{sec:discussion}.

\subsection{Flux density}
\label{sec:fluxdensitydata}

Studying the flux densities and the possible variability is one of the main goals of this paper. We have studied both the peak and the integrated flux density, obtained directly from the catalogs, for all frequencies. We were able to obtain a flux density, peak and/or integrated, measurement for 1699 out of 3998 sources, which is $42\%$ of our sources. The vast majority of the detections are from LoTSS. The hits and detection rate for each of the different frequencies can be found in Table~\ref{tab:detection}. 

\begin{figure*}
\centering
\adjustbox{valign=t}{\begin{minipage}{0.49\textwidth}
\centering
\includegraphics[width=1\textwidth]{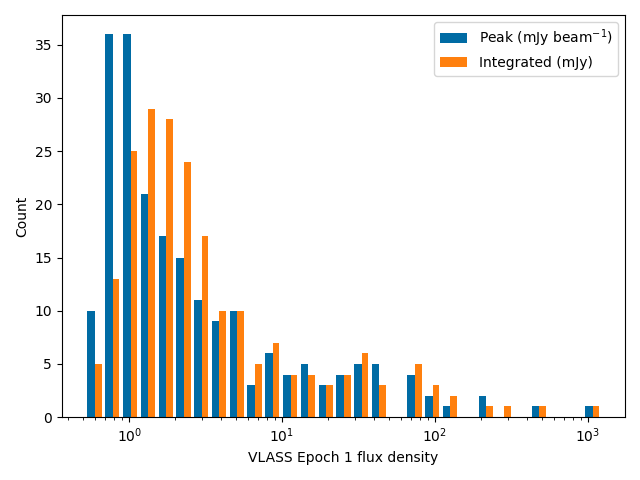}
\end{minipage}}
\adjustbox{valign=t}{\begin{minipage}{0.49\textwidth}
\centering
\includegraphics[width=1\textwidth]{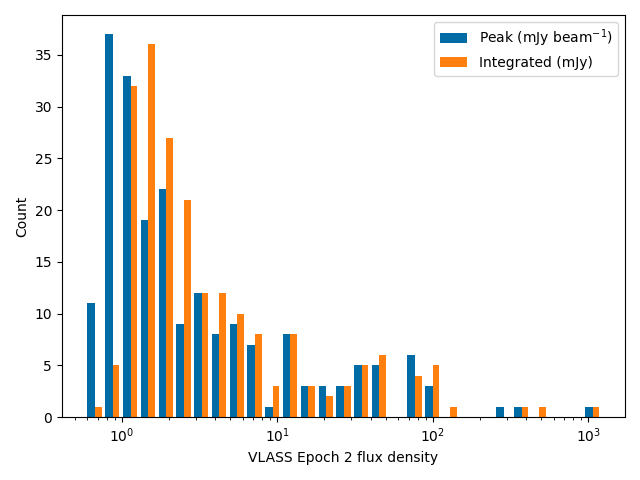}
\end{minipage}}
\hfill
\centering
\adjustbox{valign=t}{\begin{minipage}{0.49\textwidth}
\centering
\includegraphics[width=1\textwidth]{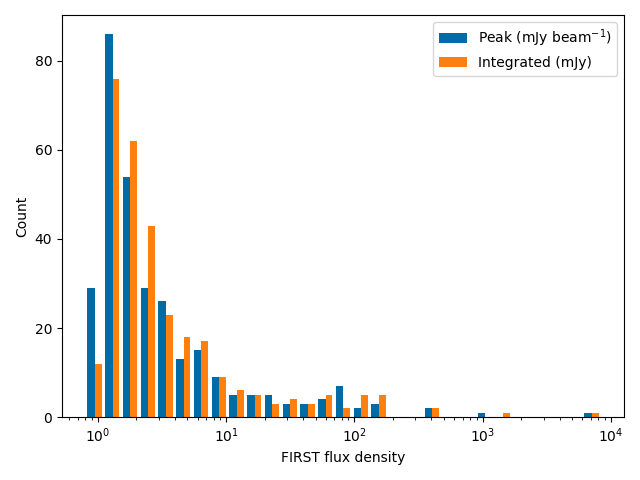}
\end{minipage}}
\adjustbox{valign=t}{\begin{minipage}{0.49\textwidth}
\centering
\includegraphics[width=1\textwidth]{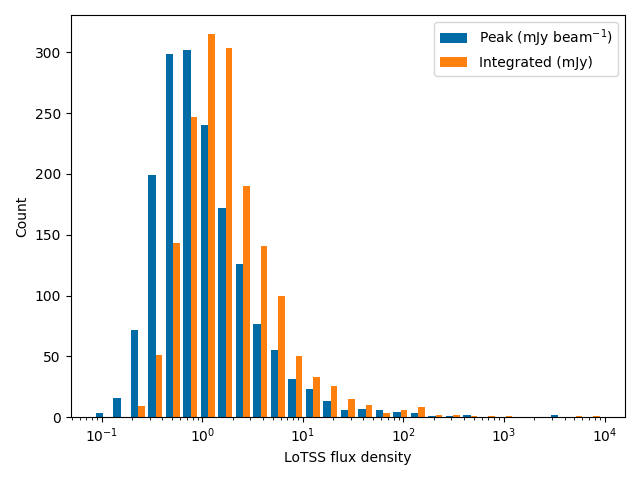}
\end{minipage}}
\hfill
    \caption{Distribution of peak and integrated flux densities. The peak flux density (mJy beam$^{-1}$) is shown in blue and the integrated flux density (mJy) is shown in orange. \emph{Top left panel:} VLASS Epoch 1, \emph{top right panel:} VLASS Epoch 2, \emph{bottom left panel:} FIRST, and \emph{bottom right panel:} LoTSS All.}  \label{fluxall}
\hfil
\end{figure*}

\begin{figure}
    \centering
    \includegraphics[width=0.49\textwidth]{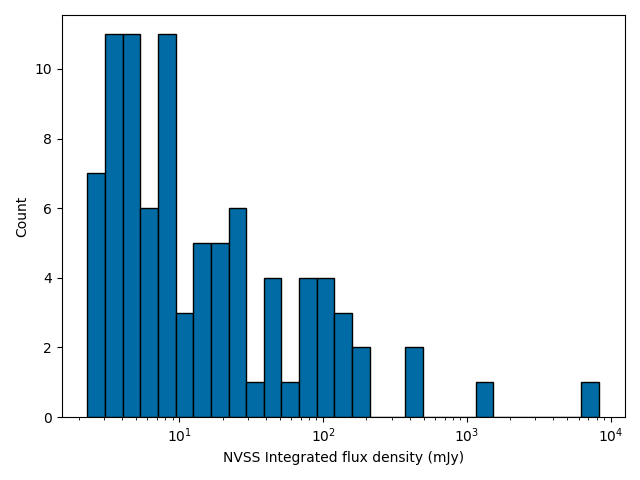}
    \caption{Distribution of NVSS Integrated flux densities in mJy.}
    \label{nvsstotflux}
\end{figure}

The key values for each survey can be found in Table~\ref{summaryflux} and will be discussed below. All flux density values for all sources can be found in the complete table, which will be uploaded to the Strasbourg Astronomical Data Centre. We have created histograms to show the flux density distribution of each survey. We present the two VLASS epochs, FIRST and LoTSS flux density histograms in Fig.~\ref{fluxall}. The NVSS histogram, seen in Fig.~\ref{nvsstotflux}, only has the integrated flux densities, as NVSS does not list the peak flux densities.

As can be seen in the histograms, a large portion of our sources have flux densities of a few mJy. However, there are some sources with very high flux densities, as is evident from Table~\ref{summaryflux}. The distributions between peak and integrated flux densities are quite similar to each other in many cases. The main differences can be noted in the fact that there are more sources in the lowest ranges for the peak flux density compared to the integrated flux densities. For some sources, the listed peak flux density is larger than the integrated flux density, however, this cannot physically occur and we have thus chosen that if $S_{\texttt{integrated}} < S_{\texttt{peak}}$ then $S_{\texttt{integrated}} = S_{\texttt{peak}}$.

\subsubsection{VLASS}

We have two sets of results for VLASS: Epoch 1 and Epoch 2. Some sources in Epoch 1 only have a peak flux density listed. The results of Epoch 2 are in line with the results of Epoch 1 for both the peak and the integrated flux densities as the flux density distributions are very similar. Nonetheless, we note that the sources detected in Epoch 1 and Epoch 2 are not identical, there are several sources only detected in one of the two epochs.

\subsubsection{FIRST}
 
The distributions of the two flux densities are very similar to each other, with the largest deviation being that there are more sources with a peak flux density of below 1~mJy. The majority of detections are in most cases very moderate, which is as expected. However, there are also some strong detections ($>100$~mJy (beam$^{-1}$)), which do not agree with the current NLS1 galaxy classification model. 

\subsubsection{LoTTS} \label{flux:lotss}

Of our surveys, LoTSS is by far the most extensive in terms of detections. Just like with FIRST, the LoTSS results are all strongly located at the lowest flux density values. There is a clear difference between the peak and integrated flux density distributions, suggesting that several sources exceed the beam size of LoTSS, thus meaning that we have several extended sources. When taking into account all LoTSS detections, the vast majority ($\sim92\%$) of sources are at least somewhat extended ($S_{\texttt{peak}}/ S_{\texttt{integrated}} \leq 0.9$). However, due to ionospheric blurring and deconvolution issues, more detailed investigation of the extendedness of the sources is not possible \citep{2022shimwell}.

\subsubsection{NVSS}

We only have the integrated flux density and its error available for NVSS. As mentioned in Sect.~\ref{sec:dataanalysis}, we prefer FIRST over NVSS due to its better resolution. There are only seven sources detected in NVSS but not FIRST. However, to showcase a comprehensive flux density distribution of NVSS, we opted to show the entire NVSS dataset. FIRST and NVSS distributions are very similar, as expected. Three of the sources detected by NVSS but not by FIRST are on the edge of the FIRST survey area. For the remaining sources, the non-detection could be due to, for example, variability and/or low surface-brightness. The flux densities of these sources are of the order of a few mJy.

\begin{table*}[t]
\caption[]{Summary of key flux density results}
\centering
\begin{tabular}{lllllllll}
Catalog                     & Min  & Max     & Mean & Median & 1. quartile & 3. quartile & Interquartile & Count \\ 
\hline\hline
FIRST Peak [mJy beam$^{-1}$]                   & 0.80 & 8073.89 & 41.93   & 1.83   & 1.31     & 4.31     & 3.00      & 302   \\
FIRST Integrated [mJy]            & 0.80 & 8240.48 & 43.48   & 2.12   & 1.43     & 4.64     & 3.21      & 302   \\
NVSS Integrated [mJy]             & 2.30 & 8283.10 & 144.60  & 8.40   & 4.40     & 32.35    & 27.95     & 88   \\
LoTSS Peak  [mJy beam$^{-1}$]             & 0.08 & 4156.34 & 8.21    & 0.82   & 0.48     & 1.67     & 1.19      & 1660  \\
LoTSS Integrated [mJy]        & 0.19 & 9267.66 & 14.23   & 1.46   & 0.85     & 2.92     & 2.07      & 1660  \\
VLASS Epoch 1 Peak [mJy beam$^{-1}$]          & 0.52 & 951.63  & 16.49   & 1.56   & 0.93     & 4.74     & 3.81      & 211   \\
VLASS Epoch 1 Integrated [mJy]    & 0.59 & 1214.08 & 19.64   & 2.08   & 1.26     & 5.42     & 4.17      & 211   \\
VLASS Epoch 2 Peak [mJy beam$^{-1}$]          & 0.58 & 971.41  & 16.64   & 1.76   & 1.01     & 5.21     & 4.20      & 207   \\
VLASS Epoch 2 Integrated [mJy]    & 0.59 & 1196.79 & 19.76   & 2.32   & 1.44     & 6.13     & 4.68      & 207   
\\\hline
\end{tabular}
\label{summaryflux}
\end{table*} 

\subsection{Variability in VLASS}
\label{sect:fracvar}

For VLASS, since we have two epochs, we were able to study changes in the flux densities. We determined the fractional variability and the absolute change. Fractional variability is a tool that is used for determining how much variability there is in a light curve. Fractional variability \citep{1992aller} is determined with:

\begin{equation}
    \Delta S = \frac{(S_\texttt{max}-\sigma_{S_\texttt{max}})-(S_\texttt{min}+\sigma_{S_\texttt{min}})}{(S_\texttt{max}-\sigma_{S_\texttt{max}})+(S_\texttt{min}+\sigma_{S_\texttt{min}})},
\end{equation}

where $S_\texttt{max}$ and $S_\texttt{min}$ denote the maximum and minimum flux densities, and $\sigma_{S_\texttt{max}}$ and $\sigma_{S_\texttt{min}}$ denote the respective errors. A positive variability index indicates that the flux density differences exceed the uncertainties while a negative result means that the uncertainties surpass the flux density difference, meaning that the source is not significantly variable. Timewise the NVSS and FIRST surveys have been completed relatively close to each other, however, due to significant resolution differences, we chose not to do the variability analysis. However, two of our sources present a considerable change between FIRST and NVSS, likely due to variability. SDSS J123355.68+130431.5 has an integrated flux density of $\sim58$~mJy in FIRST and $\sim30$~mJy in NVSS and SDSS J160048.75+165724.4 has an integrated flux density of $\sim30$~mJy in FIRST and $\sim17$~mJy in NVSS.

\begin{figure}
    \centering
    \includegraphics[width=1\linewidth]{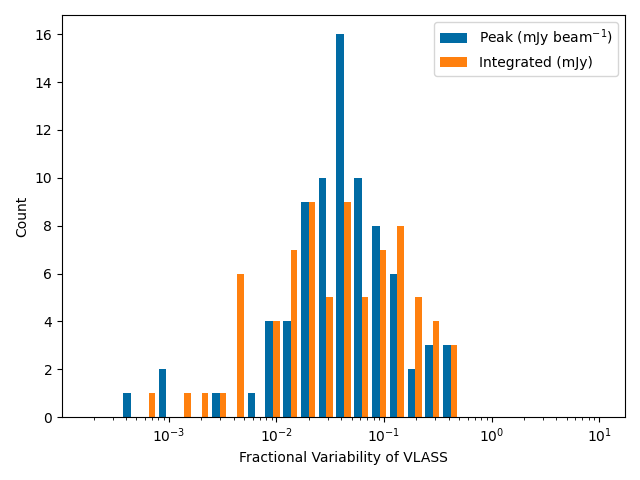}
    \caption{Fractional variability of VLASS}
    \label{fig:fracvar}
\end{figure}

We determined the fractional variability and the absolute change in the flux density between the two epochs of VLASS. There are 171 sources that have a flux density detection in both VLASS epochs. This further implies that there are 36 sources that are detected in Epoch 2 and not in Epoch 1 and 40 sources that are in Epoch 1 and not in Epoch 2. The positive variability results can be seen in Fig.~\ref{fig:fracvar}. It is clear that the vast majority of sources either have no detectable variability or very minor variability. When using the integrated flux densities, 94/171 sources have a negative variability index, and thus not detectable variability. Accordingly, 77 sources exhibit a positive variability index which means that these sources do have detectable variability. However, a low fractional variability can be a result of the flux uncertainties, which is why we implement a threshold of 0.1 for the fractional variability index. Of the 77 sources, 20 have a fractional variability index $>$ 0.1. The maximum variability index for the integrated flux density results is 0.38. When we only take into account sources with a positive variability index, the mean index is $\sim0.08$ and the median is $\sim0.04$. If we account for the VLASS flux density uncertainties, there are 79 sources with a positive fractional variability index and 92 sources with a negative fractional variability index for the integrated flux density results. However, the number of sources with a fractional variability index of more than 0.1 stays the same: 20 sources. 

For the peak flux density, there are 91 sources that have a negative variability index. Detectable variability is seen in 80 sources. The maximum variability index is $\sim0.38$. The mean variability index for the sources with a detectable variability is $\sim0.07$ and the respective median value is $\sim0.04$. When taking into account the flux density uncertainties, a positive fractional variability index can be found in 81 sources and a negative index in 90 sources. The number of sources with a fractional variability index of more than 0.1 increases in this case from 16 to 17. All in all, the flux density underestimations do not noticeably change the results.

If we take into account sources that are only detected in one of the two epochs and determine the 5$\sigma$ upper limit of the flux densities of these sources, the number of variable sources in VLASS increases. In this case, there are 247 sources for which the fractional variability can be determined. For the integrated flux density results the number of detectably variable sources increases to 136 sources and for the peak flux density results to 116 sources. Out of the 136 sources, 54 have a fractional variability index above 0.1. For the peak flux density results the corresponding number of sources is 28. These sources are not included in Fig.~\ref{fig:fracvar}. We determined the upper limit of the flux density with the median rms of the VLASS Quick Look catalog.

For the absolute change of the integrated flux densities, we subtract Epoch 2 values from Epoch 1 values. The maximum positive absolute change is 122.69~mJy, and the maximum negative absolute change is -35.42~mJy. The median absolute change is -0.09~mJy and the mean is 0.65~mJy. Both of these values are small, suggesting that the majority of the sources do not experience significant variability between the epochs. For the peak flux density, the absolute change results are quite similar to those of the integrated flux densities. The largest positive absolute change is 132.97~mJy beam$^{-1}$. The largest negative absolute change is -65.02~mJy beam$^{-1}$. The mean and median absolute changes are 0.50~mJy beam$^{-1}$ and -0.06~mJy beam$^{-1}$, respectively.

\subsection{Spectral index}
\label{dataspecind}

The spectral index can tell us about the origin and the production mechanism of the radio emission. To determine the spectral index, we follow the convention $ S_\nu \propto \nu^\alpha$, where $S$ is the emitted flux density, $\nu$ denotes the frequency, and $\alpha$ the power-law index. To calculate the spectral index we used

\begin{equation}
    \alpha_{\texttt{lf,hf}} = \frac{\log{S_\nu^\texttt{hf}}-\log{S_\nu^\texttt{lf}}}{\log{\nu^\texttt{hf}}-\log{\nu^\texttt{lf}}},
\end{equation}

where $\alpha$ is the spectral index, hf and lf stand for higher and lower frequency respectively, and $S$ is the peak or integrated flux. As mentioned in Sect.~\ref{dataspecind}, we study both flux density types for the spectral index. This is done due to the total and peak flux densities providing clues on different aspects. We note that at a higher resolution, it is possible for some of the flux to be resolved out when compared to the lower frequencies which can cause the spectral index to appear steeper than it is. However, we do not expect this to significantly change the results. We have determined the spectral index for the pairs: VLASS and FIRST, and VLASS and LoTSS. This has been done for both epochs of VLASS. The central frequencies in our calculations were 3~GHz, 1.4~GHz, and 144~MHz for VLASS, FIRST, and LoTSS, respectively. We have also determined the error of the spectral index using the formula

\begin{equation}
    \Delta\alpha=\frac{\log_{10}e}{\log_{10}\frac{\nu_{\texttt{hf}}}{\nu_{\texttt{lf}}}} \left[ \frac{\Delta S_{{\texttt{hf}}}}{S_{\texttt{hf}}} - \frac{\Delta S_{{\texttt{lf}}}}{S_{\texttt{lf}}}\right ] .
\end{equation}

Observationally AGN spectra can be divided into three types: inverted, flat, and steep. An inverted spectrum is defined as having $\alpha >$ 0.5, whereas a steep-spectrum source has $\alpha <$ $-$0.5. Thus a source with a spectral index between these two types ($-$0.5 $< \alpha < 0.5$) has a flat spectrum. Radio emission from star formation usually has two components, free-free emission from ionized gas and synchrotron radiation from supernova remnants. The characteristic spectral indices for free-free emission and supernova remnants are $-$0.1, and $-$0.8 to $-$1.0, respectively \citep{1998condon}. In the case of these two coexisting, which often occurs when there is star formation, a superposition gives a resultant spectral index of $\sim-0.7$. This result is near the optically thin synchrotron emission from AGN, making the interpretation of spectral indices close to this value challenging, especially if the information on the radio morphology is lacking. As we work at low frequencies, we likely start seeing the effects of free-free absorption and/or synchrotron self-absorption. This would manifest as lower-than-expected low-frequency flux density values when extrapolating from higher frequencies. However, with these data, it will be hard to study this as the data are not simultaneous. In most cases, a flat or an inverted spectral index is an indication that the predominant source of the radio emission is the AGN. This emission usually originates from the optically thick radio core or the jet, especially when seen at small angles. In the case of the jets being kinematically young and synchrotron self-absorbed, the resulting spectrum peaks in the MHz/GHz range. Under ideal conditions, the maximum spectral index of synchrotron self-absorbed plasma is +2.5, but it is often observed to be shallower than this. An even higher spectral index is linked with free-free absorption.

\subsubsection{LoTSS v. FIRST and FIRST v. VLASS spectral indices}

\begin{figure*}
\centering
\adjustbox{valign=t}{\begin{minipage}{0.49\textwidth}
\centering
\includegraphics[width=1\textwidth]{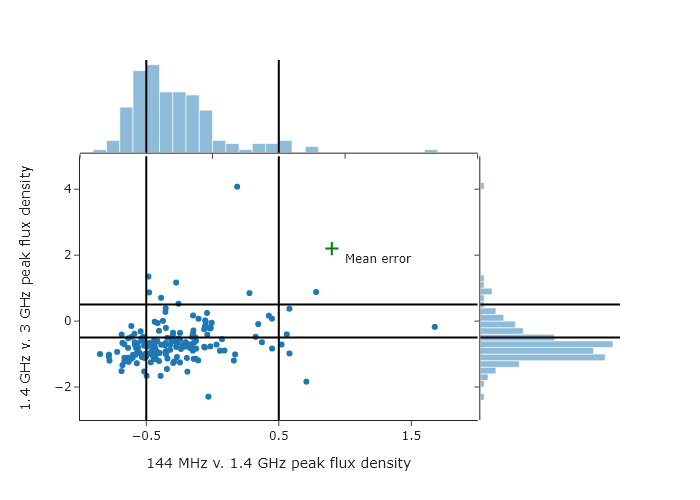}
\end{minipage}}
\adjustbox{valign=t}{\begin{minipage}{0.49\textwidth}
\centering
\includegraphics[width=1\textwidth]{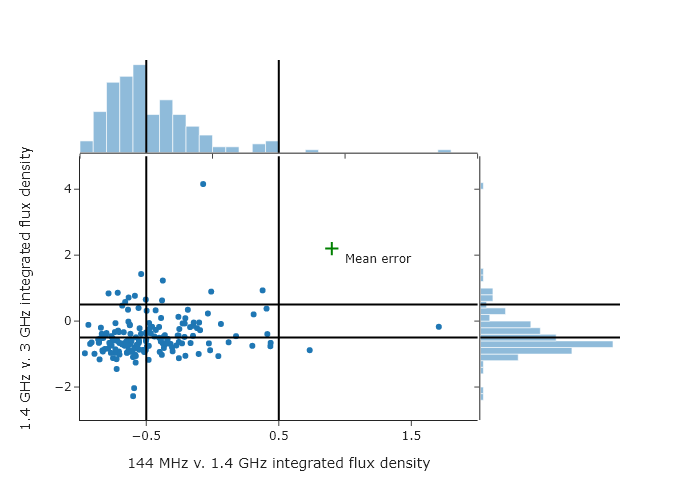}
\end{minipage}}
\hfill
    \caption{\emph{Left panel:} Spectral index plot using 144~MHz and 1.4~GHz v. 1.4~GHz and 3~GHz (Epoch 1), peak flux densities. \emph{right panel:} Spectral index plot using 144~MHz and 1.4~GHz v. 1.4~GHz and 3~GHz (Epoch 1), integrated flux densities. The average error is plotted with green.
    }  \label{alpha1}
\hfil
\end{figure*}

\begin{figure*}
\centering
\adjustbox{valign=t}{\begin{minipage}{0.49\textwidth}
\centering
\includegraphics[width=1\textwidth]{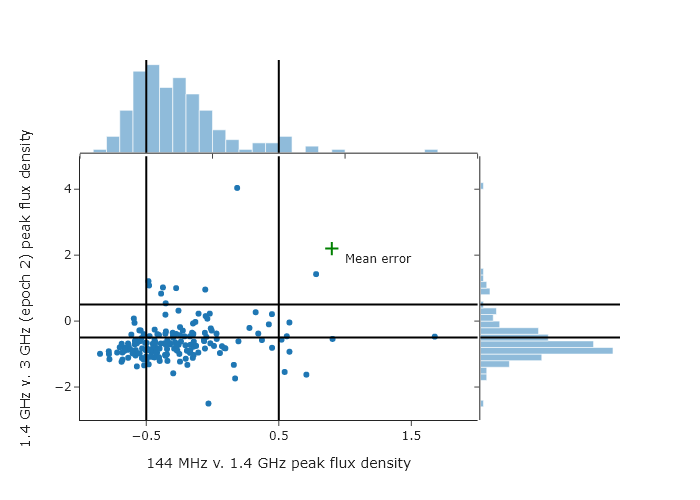}
\end{minipage}}
\adjustbox{valign=t}{\begin{minipage}{0.49\textwidth}
\centering
\includegraphics[width=1\textwidth]{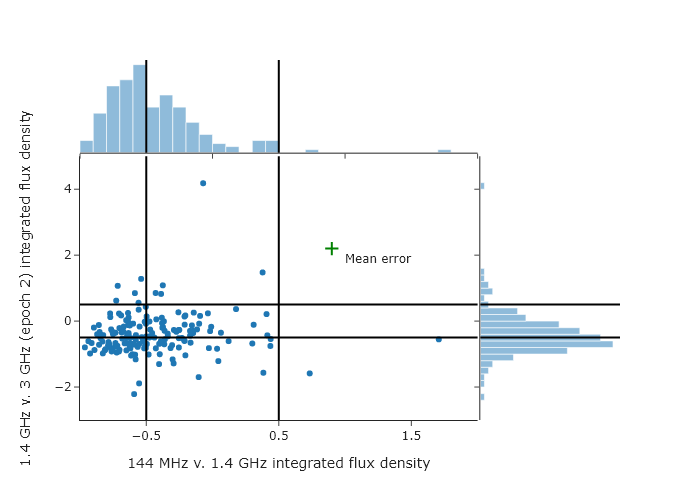}
\end{minipage}}
\hfill
    \caption{\emph{Left panel:} Spectral index plot using 144~MHz and 1.4~GHz v. 1.4~GHz and 3~GHz (Epoch 2), peak flux densities. \emph{right panel:} Spectral index plot using 144~MHz and 1.4~GHz v. 1.4~GHz and 3~GHz (Epoch 2), integrated flux densities. The average error is plotted with green.
    }  \label{alpha2}
\hfil
\end{figure*}

We have two alpha-alpha plots for the following pairs: 144~MHz and 1.4~GHz (LoTSS and FIRST), and 1.4~GHz and 3~GHz (FIRST and VLASS Epoch 1 or VLASS Epoch 2). These plots can be seen in Fig.~\ref{alpha1} and Fig.~\ref{alpha2} and only include sources that have been detected at all three frequencies. To improve readability, we added lines at $-0.5$ and 0.5 for both axes in the plots. We present results using both peak and integrated flux densities, as they give information on different properties of the source. For extended sources, the peak flux density result can provide information on the AGN, whereas the integrated flux density can tell us more about the nature of the extended emission of the source. The number of sources that have an integrated flux density larger than the peak flux density by at least $\sim10\%$ is 108, 124, and 71, for VLASS Epoch 1, VLASS Epoch 2, and FIRST, respectively. A summary table, Table~\ref{summaryspind}, has all the key value information on the spectral indices. All spectral index results and their implications will be discussed in more detail in Sect.~\ref{radioemission}.

When studying Fig.~\ref{alpha1} and Fig.~\ref{alpha2}, sources in the bottom left corner have a steep spectrum both for 144~MHz v. 1.4~GHz and 1.4~GHz v. 3~GHz, suggesting that the radio emission is most likely caused by either star formation or optically thin AGN emission. Sources located in the bottom right corner are interesting, as they have an increasing spectrum from 144~MHz to 1.4~GHz and a decreasing spectrum from 1.4~GHz to 3~GHz, thus making them peaked source candidates. There are three clear peaked source candidates. There are also a few sources for which one of the spectral indices is very extreme, likely indicating significant variability.

We were able to calculate the 144~MHz and 1.4~GHz spectral index for 232 sources using both peak and integrated flux densities. For the peak flux density spectral index results, there are 161 sources with a flat spectrum, 66 with a steep spectrum, and five with an inverted spectrum. For the integrated flux density spectral index results, the respective results are 95 sources with a flat spectrum, 136 with a steep spectrum, and one with an inverted spectrum. Some of the most interesting individual sources are those with either very inverted or very steep spectra as cases of very high spectral indices are often associated with significant AGN activity, causing variability. These will be discussed in Sect.~\ref{radioemission}.

We have a spectral index result for 189 sources using the peak flux densities and for 187 sources using integrated flux densities when using VLASS Epoch 1 and FIRST. Of the 189 sources, 132 have a spectral index result below $-0.5$, which means that $\sim71\%$ of these sources have a steep spectrum. There are 46 sources that have a flat spectral index, and nine of our sources have an inverted spectrum. If we take into account the flux underestimations of VLASS, the results change slightly, the number of steep spectrum sources decreases to 120 whereas the number of flat spectrum sources increases to 58. The number of inverted sources remains the same. There are several steep-spectrum outliers ($\alpha < -1.25$). One of the most clear outliers is the very inverted spectrum of SDSS J110546.06+145202.4 with $\alpha \sim$ 4.08. Such an inverted spectrum could be a result of variability or in the case of no variability, the result could suggest free-free absorption (FFA). The spectral index results when using the integrated flux density, there are 108 sources that classify as steep-spectrum sources, which is significantly less than for the peak flux density results. Of the 187 sources, 66 sources present flat spectra. The final 13 sources show an inverted spectrum. When taking into account the flux density uncertainties, there are 104 steep- sources, 69 flat- sources, and 14 inverted-spectrum sources.

The VLASS Epoch 2 and FIRST results are very similar to those of VLASS Epoch 1 and FIRST. For the peak flux density results, there are 129 source with a steep spectrum, 50 with a flat spectrum, and ten with an inverted spectrum. For the integrated flux density sources, there are 89 sources with a flat spectrum, 90 with a steep spectrum, and still ten with an inverted spectrum. The results when considering the flux uncertainties change in a similar manner as they did for the VLASS Epoch 1 v. FIRST. For the peak flux density results, the number of sources with a steep spectrum decreases to 115 and the number of flat spectrum sources increases to 64. The number of sources with an inverted spectrum remains the same. For the integrated flux density results, there are 83 steep spectrum sources, 95 flat spectrum sources, and 11 inverted spectrum sources.

\subsubsection{Variability and spectral index}

\begin{figure*}
\centering
\adjustbox{valign=t}{\begin{minipage}{0.49\textwidth}
\centering
\includegraphics[width=1\textwidth]{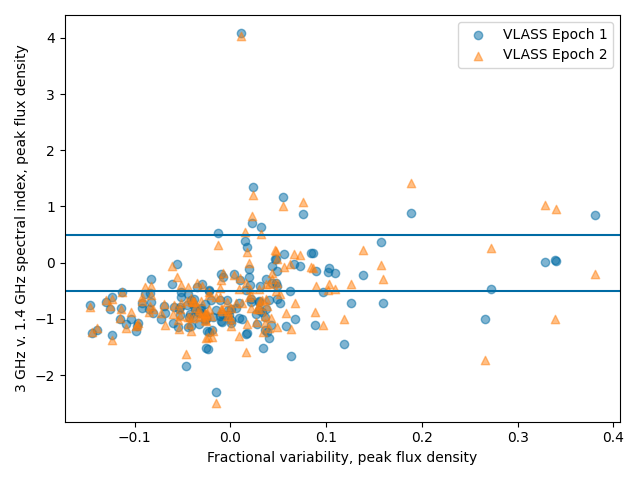}
\end{minipage}}
\adjustbox{valign=t}{\begin{minipage}{0.49\textwidth}
\centering
\includegraphics[width=1\textwidth]{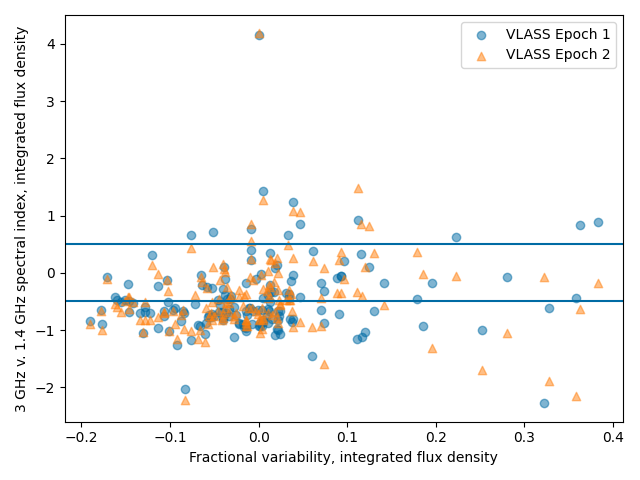}
\end{minipage}}
\hfill
    \caption{Relation of spectral index and fractional variability. \emph{Left panel:} Peak flux density results and \emph{right panel:} integrated flux density results. Blue lines have been added at the limits for a steep spectrum ($\alpha = -0.5$) and an inverted spectrum ($\alpha = 0.5$)}. \label{varsi}
\hfil
\end{figure*}

Studying the relationship between the fractional variability index and the spectral index, shown in Fig~\ref{varsi}, we can see that the higher variability indices ($>0.2$) occur for all of the different spectral types (steep, flat, and inverted). However, the majority of the non-variable sources have a steep spectrum. We note that as our data is not simultaneous it may affect the distribution between the various spectral types. For the integrated flux density results, there are several sources that have very steep spectral indices ($<-1$). Such steep spectral indices can indicate significant radio variability. The source with the most inverted spectrum does not showcase clear variability between the two VLASS epochs. All in all, the spectral index results of this study are inline with the results of prior studies \citep[e.g.,][]{2010foschini, 2016berton, 2018berton, 2021jarvela, 2024jarvela}.

\begin{table*}
\caption{Summary of key spectral index results}
\centering
\begin{tabular}{lllllllll}
Catalog                           & Min   & Max  & Mean & Median & 1. quartile & 3. quartile & Interquartile & Count \\ 
\hline\hline
FIRST v. LoTSS peak [mJy beam$^{-1}$]          & -0.88 & 1.68 & -0.30   & -0.35  & -0.51   & -0.17  & 0.34    & 269   \\
FIRST v. LoTSS integrated [mJy]            & -1.03 & 1.71 & -0.46   & -0.51  & -0.67   & -0.30  & 0.37    & 269   \\
VLASS Epoch 1 v. FIRST peak [mJy beam$^{-1}$]  & -2.29 & 4.08 & -0.65   & -0.76  & -1.01   & -0.42  & 0.58    & 187   \\
VLASS Epoch 1 v. FIRST integrated [mJy]    & -2.28 & 4.16 & -0.46   & -0.61  & -0.81   & -0.21  & 0.60    & 187   \\
VLASS Epoch 2 v. FIRST peak [mJy beam$^{-1}$]  & -2.50 & 4.03 & -0.61   & -0.71  & -0.95   & -0.40  & 0.56    & 189   \\
VLASS Epoch 2 v. FIRST integrated [mJy]    & -2.21 & 4.18 & -0.41   & -0.47  & -0.76   & -0.13  & 0.63    & 189   

\\\hline
\end{tabular}
\label{summaryspind}
\end{table*}

\subsection{Radio luminosity}

Radio luminosity is often used as a proxy for the radio activity of the AGN. Radio luminosity has a key advantage compared to the spectral index: only one frequency is required thus eliminating the need for synchronous data. However, it is important to note that NLS1 galaxies are a peculiar bunch and using the radio luminosity as an indicator for radio activity is not always suitable; although a high radio luminosity is often indicative of AGN-produced radio emission, in NLS1 galaxies, a low radio luminosity does not necessarily mean a lack of AGN-produced radio emission. On average star-forming and starburst galaxies generally have $\log_{10}(\nu L_{\nu})<40$ erg $s^{-1}$ \citep{2009sargsyan}, determined at 1.4~GHz. For the other frequencies in our study, the threshold is $\log_{10}(\nu L_{\nu})>40.1$ erg s$^{-1}$ for 3~GHz and $\log_{10}(\nu L_{\nu})>39.7$ erg s$^{-1}$  for 144~MHz. To keep these results consistent with the analysis of \citet{2009sargsyan}, we have estimated these values with the same spectral index $-0.7$.

We determined the k-corrected radio luminosity using
\begin{equation}
    \nu L_\nu = \frac{4\pi \nu d^2_{L} S_{\nu}}{(1+z)^{({1-\alpha})}} \quad [\mathrm{erg\,s^{-1}}],
\end{equation}
where $\nu$ is the central frequency in Hz, $d_L$ is the luminosity distance in cm, $S_\nu$ is the flux density at the central frequency in erg s$^{-1}$ cm$^{-2}$ Hz $^{-1}$, and $z$ is the redshift. The luminosity distance was determined by using CosmoCalc \citep{2006wright}. For the sake of simplicity, we have chosen to use a fixed spectral index value of $\alpha = -0.5$. The value was chosen as it is the break point between a steep and a flat spectrum. Furthermore, by using a fixed value we avoid any possible issues with the calculated spectral indices as none of our data is synchronous. Finally, the spectral index will not cause any large discrepancies even if it would change significantly. We have determined the radio luminosity for FIRST, VLASS (both epochs), and LoTSS. We have also studied the extendedness of the sources compared to the radio luminosity. 

As with the spectral index, high radio luminosity typically indicates significant AGN activity. As we make use of a wide range of frequencies, we quote luminosities in terms of $\nu L_\nu$ so that they are more easily comparable. The logarithmic radio luminosity distributions VLASS (both epochs), FIRST, and LoTSS using both peak and integrated flux densities are shown in Fig.~\ref{lumfdrboth} and Fig.~\ref{lume12both}. The radio luminosities of NVSS can be found in the complete data table. We are aware that the radio luminosity is generally only calculated for the integrated flux densities, but we decided to also calculate it using the peak flux densities to improve the chances of identifying curious behavior. A summary table, Table~\ref{table:radiolumquartile}, can be found in Sect.~\ref{sec:discussion} with the key values of the different radio luminosities. 

\subsubsection{FIRST}

The distribution of radio luminosity results of both peak and integrated, seen in Fig.~\ref{lumfdrboth}, flux densities are very similar. There are 35 sources that have a higher radio luminosity than $10^{41}$ erg s$^{-1}$, this holds true for both the peak and integrated results. Although a lot of the sources have moderate radio luminosities, nearly half of the sources, $\sim42\%$ and $\sim44\%$ for peak and integrated flux density results, respectively, have a radio luminosity exceeding $10^{40}$ erg s$^{-1}$.

\subsubsection{LoTSS}

\begin{figure*}
\centering
\adjustbox{valign=t}{\begin{minipage}{0.49\textwidth}
\centering
\includegraphics[width=1\textwidth]{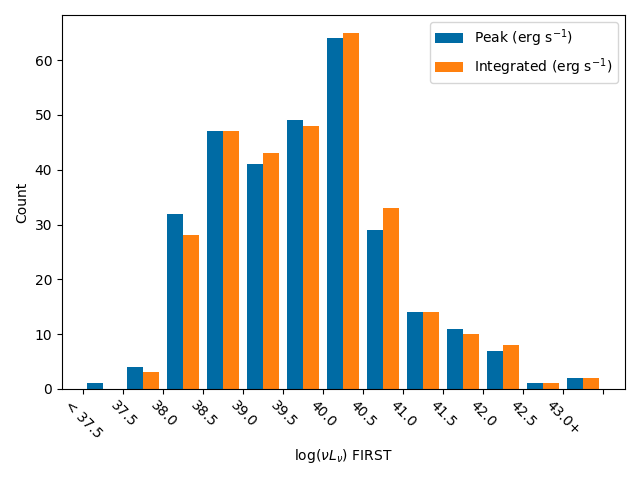}
\end{minipage}}
\adjustbox{valign=t}{\begin{minipage}{0.49\textwidth}
\centering
\includegraphics[width=1\textwidth]{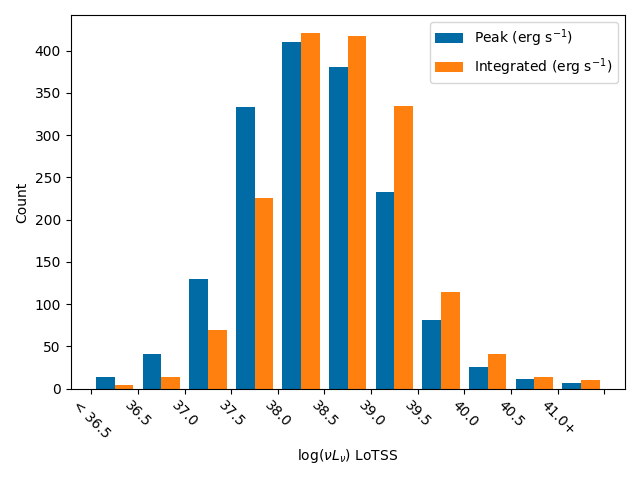}
\end{minipage}}
\hfill
    \caption{Logarithmic radio luminosity distribution of peak and integrated flux densities. Peak flux density results are shown in blue and integrated flux density results are shown in orange. \emph{Left panel:} FIRST and \emph{right panel:} LoTSS All.}  \label{lumfdrboth}
\hfil
\end{figure*}

The radio luminosity distributions results can be seen in Fig.~\ref{lumfdrboth} for peak and integrated flux densities. The distribution shows very few extremely luminous or faint sources. There are some sources, $\sim5\%$ and $\sim8\%$ for peak and integrated flux density results respectively, with a radio luminosity above $10^{39.7}$ erg s$^{-1}$, which we take to be the threshold for significant AGN activity at 144~MHz. This result was expected as the radio emission from star formation is significant at 144~MHz.

\subsubsection{VLASS}

\begin{figure*}
\centering
\adjustbox{valign=t}{\begin{minipage}{0.49\textwidth}
\centering
\includegraphics[width=1\textwidth]{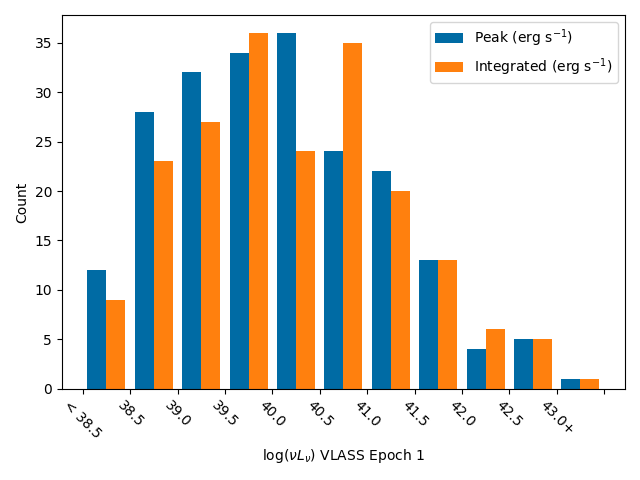}
\end{minipage}}
\adjustbox{valign=t}{\begin{minipage}{0.49\textwidth}
\centering
\includegraphics[width=1\textwidth]{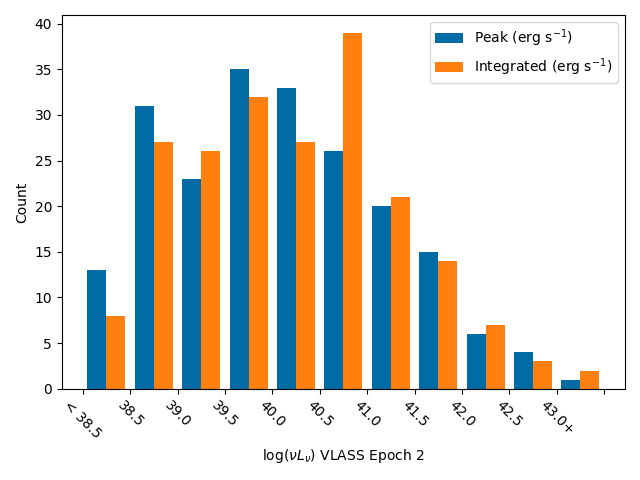}
\end{minipage}}
\hfill
    \caption{Logarithmic radio luminosity distribution of peak and integrated flux densities. Peak flux density results are shown in blue and integrated flux density results are shown in orange. \emph{Left panel:} VLASS Epoch 1 and \emph{right panel:} VLASS Epoch 2.}  \label{lume12both}
\hfil
\end{figure*}

The VLASS data consistently show the highest radio luminosities. The distributions of the peak and integrated flux density radio luminosity of Epoch 1 and Epoch 2, seen in Fig.\ref{lume12both}, have only minor differences. For both Epoch 1 and Epoch 2, peak and integrated results, roughly half of the sources have a radio luminosity above $10^{40}$ erg s$^{-1}$, with $\sim5-6\%$ of sources presenting radio luminosities above $10^{42}$ erg s$^{-1}$. These fractions remain the same even when taking into account the flux density underestimations.

\subsubsection{Variability and radio luminosity}

\begin{figure*}
\centering
\adjustbox{valign=t}{\begin{minipage}{0.49\textwidth}
\centering
\includegraphics[width=1\textwidth]{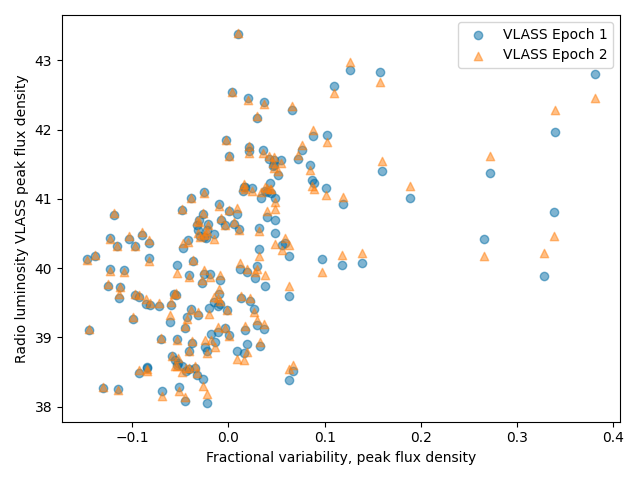}
\end{minipage}}
\adjustbox{valign=t}{\begin{minipage}{0.49\textwidth}
\centering
\includegraphics[width=1\textwidth]{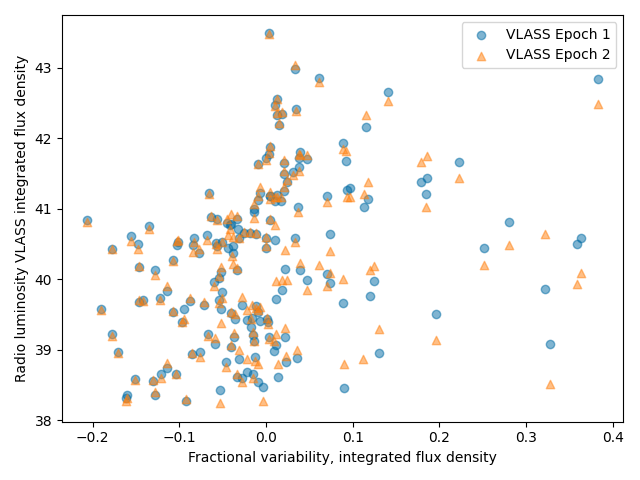}
\end{minipage}}
\hfill
    \caption{Relation of the logarithmic radio luminosity and fractional variability. \emph{Left panel:} Peak flux density results and \emph{right panel:} integrated flux density results.}  \label{varrl}
\hfil
\end{figure*}

When studying the relation between the fractional variability and the radio luminosity, seen in Fig.\ref{varrl}, it can be seen that the higher variability indices, $>0.2$, generally occur in sources with higher radio luminosities $10^{40}$ erg s$^{-1}$. Given that the radio luminosities are above $10^{40}$ erg s$^{-1}$, the AGN is the most likely culprit for this variability behavior.

\subsection{Relation of extendedness to radio luminosity}

We studied the relationship between the extendedness and radio luminosity of VLASS Epoch 1 and 2, and FIRST. In this paper, we are determining the extendedness by dividing the peak flux density by the integrated flux density. If the resulting quotient is less than 0.9, then the source is deemed extended. We note that for low flux densities, a noise uncertainty can cause a source to appear extended even if it is not. We have added a line from the point where the quotient is at 0.9 as help for visualization to easily see if a source is extended or point-like. 

\begin{figure*}
\centering
\adjustbox{valign=t}{\begin{minipage}{0.49\textwidth}
\centering
\includegraphics[width=1\textwidth]{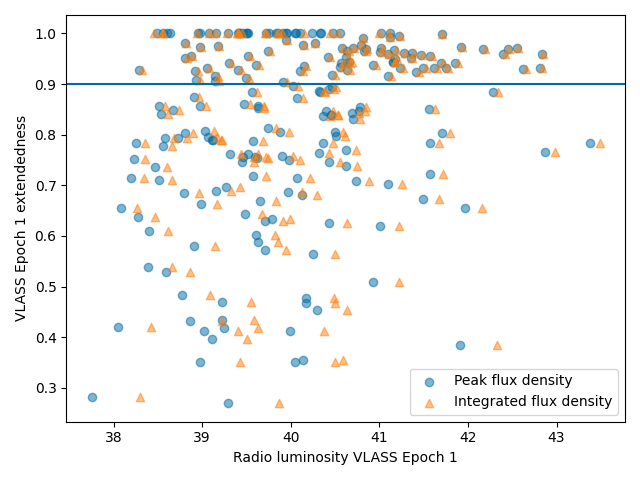}
\end{minipage}}
\adjustbox{valign=t}{\begin{minipage}{0.49\textwidth}
\centering
\includegraphics[width=1\textwidth]{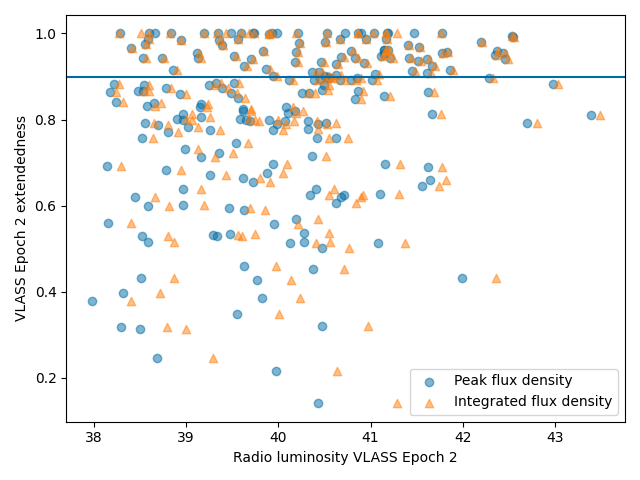}
\end{minipage}}
\adjustbox{valign=t}{\begin{minipage}{0.49\textwidth}
\centering
\includegraphics[width=1\textwidth]{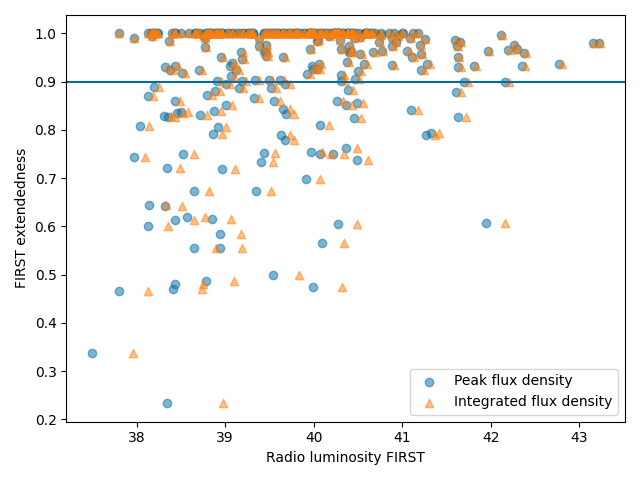}
\end{minipage}}
\hfill
    \caption{Relation of the extendedness compared to the logarithmic radio luminosity. Radio luminosity is given on the x-axis and the extendedness on the y-axis. The line at an extendedness of 0.9 is added to show possible extendedness. \emph{Top left panel:} VLASS Epoch 1, \emph{top right panel:} VLASS Epoch 2, and \emph{bottom panel:} FIRST.}  \label{comprl}
\hfill
\end{figure*}

The VLASS Epoch 1, VLASS Epoch 2, and FIRST results can be seen in Fig.~\ref{comprl}. For the VLASS Epoch 1 results, 108 sources ($\sim51\%$) appear extended. For VLASS Epoch 2 there are 124 sources ($\sim60\%$) that satisfy the used extendedness criterion. For FIRST the respective value is 71 sources out of 302 ($\sim24\%$). When we study the VLASS results, we can see that the extendedness is more dispersed for the lower radio luminosities. However, this could partially be caused by noise for fainter sources. Noise is most likely not the culprit for all of the dispersion, as the errors at low flux densities ($S < 1$~mJy), are rather small when compared to the respective flux densities. Furthermore, with the used extendedness criteria, the effects of noise are not as strong. Sources with higher radio luminosities seem to have a higher tendency to be compact. From the plot, it is possible to see that several sources are very compact, with the extendedness being close to 1. There are, however, also a significant amount of sources that appear extended. A few sources with the highest radio luminosities, $>10^{42}$ erg s$^{-1}$, do seem to present clearly extended structure in the VLASS plots, but not in FIRST. This is likely due to the different resolutions of the surveys. Although by visual inspection there are a few genuinely extended sources in the LoTSS images, it is not possible to analyze the LoTSS data in terms of peak to extended flux ratio as this is known to be unreliable in LoTSS for faint sources due to ionospheric blurring and deconvolution issues \citep{2022shimwell}. Accordingly we do not apply this analysis to the LoTSS data.

\begin{table*}[]
\caption{Summary of key radio luminosity results}
\centering
\begin{tabular}{lllllllll}
Quartile results               & Min      & Max      & Mean  & Median & 1. quartile & 3. quartile & Interquartile & Count \\
\hline\hline
VLASS Epoch 1 Peak [mJy beam$^{-1}$]       & 37.75 & 43.38 & 40.06   & 39.99  & 39.20  & 40.75     & 1.55          & 211   \\
VLASS Epoch 1 Integrated [mJy] & 38.27 & 43.49 & 40.19   & 40.13  & 39.36  & 40.85     & 1.49          & 199   \\
VLASS Epoch 2 Peak [mJy beam$^{-1}$]       & 37.98 & 43.39 & 40.09   & 40.08  & 39.22  & 40.84     & 1.62          & 207   \\
VLASS Epoch 2 Integrated [mJy] & 38.24 & 43.48 & 40.19   & 40.19  & 39.31  & 40.92     & 1.61          & 206   \\
FIRST Peak [mJy beam$^{-1}$]               & 37.49 & 43.22 & 39.78   & 39.79  & 38.93  & 40.40     & 1.48          & 302   \\
FIRST Integrated [mJy]         & 37.80 & 43.23 & 39.82   & 39.80  & 38.97  & 40.43     & 1.46          & 302   \\
LoTSS Peak [mJy beam$^{-1}$]               & 35.62 & 42.70 & 38.42   & 38.39  & 37.88  & 38.92     & 1.04          & 1660  \\
LoTSS Integrated [mJy]         & 35.45 & 42.80 & 38.65   & 38.60  & 38.13  & 39.12     & 0.99          & 1660  
\\\hline
\end{tabular}
\tablefoot{All of the results are given in the logarithmic form.}
\label{table:radiolumquartile}
\end{table*}

\section{Optical data analysis and results}
\label{optspec}

The optical spectral properties are what currently define NLS1 galaxies. For this paper, we wanted to study whether or not there is a connection between any of the spectral properties and the flux density, spectral index, and/or radio luminosity. The common spectral properties to study in NLS1 galaxies include the flux of the broad and narrow $H\beta$ components (F(H$\beta$b) and F(H$\beta$n), respectively), the flux of the forbidden doubly ionized oxygen line (F[O III]), the FWHM of the broad $H\beta$ line (FWHMH($\beta$)), the optical Fe II strength relative to the broad $H\beta$ component in the range of 4434~Å to 4684~Å (R4570), the monochromatic luminosity at 5100 Å (L5100), and the Eddington ratio. We have opted not to study the F(H$\beta$n) as the [O III] essentially provides the same information. We also studied how the redshift, black hole mass, and Eddington ratio compare to the listed parameters. The spectral properties used in this paper are from \citet{2017rakshit}. When modeling the spectra \citet{2017rakshit} have used a single local power-law continuum component as well as a template from \citet{2010kovacevic} for Fe II. Two Gaussian functions have been used for modeling the wavelength range of 4385 Å to 5500 Å, whereas for the region 6280 Å to 6750 Å only one Gaussian function is used. All the components have been modeled simultaneously, using a fixed theoretical value of three for the flux ratio of both [O III] and [N II] doublets. For redshifts higher than 0.3629, only the H$\beta$ was modeled. Furthermore, \citet{2017rakshit} has fitted a single Lorentzian and a single Gaussian function on both H$\beta$ and H$\alpha$ lines. The better fit was determined by the Chi-squared test. No S/N ratio criterion was used by \citet{2017rakshit} when finalizing the sample set. In the future, the results of our study need to be confirmed with a more detailed study using higher quality optical spectral data, employing more sophisticated modeling of the data and statistical methods, such as a self-organizing map study or principal component analysis study. As mentioned in Sect.~\ref{sec:sample}, although the spectral measurements in \citet{2017rakshit} are not ideal, due to the cleaning process carried out in \citet{2020berton2}, the spectral results for this sample should be of higher-than-average quality in comparison to the entire sample of \citet{2017rakshit}.

\subsection{Optical spectral property results}
\label{specresults}

The key interesting optical spectral properties are as mentioned in Sect.~\ref{optspec}. There are several known relations between the mentioned parameters and redshift, black hole mass, and Eddington ratio: these relations will be noted, but not analyzed further, in this paper. The deepest analysis will be on the extreme behavior and the possible relations to other observables in that aspect. All figures relating to this can be found in the Appendix~\ref{appextreme}.

The extreme behaviors that we are studying are on: radio luminosity, fractional variability, absolute change, spectral index, and extendedness. We are taking into account both extreme ends for the spectral index, the radio luminosity, and the absolute change. We are only using the negative extreme end for the extendedness parameters and only the positive extreme end for the fractional variability property. We have used varying sigmas to get the ends in the analysis. The variability label is given for sources that have extreme fractional variability and/or extreme absolute change. The used sigmas for each property as well as the number of sources can be found in the Appendix~\ref{appextreme} in Table~\ref{tab:extreme}. In the optical spectral plots, we have marked all sources that have any type of extreme behavior. In the cases of a source having multiple extreme behaviors, the symbols for the extreme types will be superimposed. The spectral results and their uncertainties can be found in \citet{2017rakshit}. 

We are using a simple Spearman's correlation rank to study the statistical relations. We note that our analysis is not an extensive statistical study and should not be considered as such, however, this will give a basic idea of the behavior. With the Spearman's correlation rank the interest is in the correlation coefficient, $r_S$, and in the p-value. The correlation coefficient ranges from -1 to 1, with negative values meaning a negative correlation and positive values indicating a positive correlation. The significance levels of the correlation coefficient and what is deemed strong correlation vary in literature, but a commonly used limits are $r_S < 0.1$ for no correlation, $0.1 \leq r_S < 0.3$ for weak correlation, $0.3 \leq r_S < 0.5$ as moderate correlation, $0.5 \leq r_S < 0.7$, as strong correlation, and finally $r_S \geq 0.7$ as very strong correlation \citep{kuckartz2013}. The p-value is related to the null hypothesis. For Spearman's correlation rank, the null hypothesis is that there is no correlation.

\textit{FWHMH($\beta$)} When comparing the FWHM(H$\beta$) to Eddington ratio, black hole mass, and redshift, seen in Fig.~\ref{fwhm}, three different relations can be seen. There is a clear anti-correlation relation between the FWHM(H$\beta$) and the Eddington ratio. This is also shown by $r_S \approx -0.58$, meaning that the anti-correlation relation is quite strong. The p-value is 0, and thus the null hypothesis is rejected. Nearly all sources ($\sim97\%$) with an Eddington ratio of more than 2 have a FWHM(H$\beta$) of less than 1000 km/s. This could be caused by the interconnections between the FWHM(H$\beta$), the black hole mass, and the Eddington ratio. Sources with FWHM(H$\beta$) less than 500 km s$^{-1}$ have black hole masses ($\log\frac{M_{BH}}{M_{\sun}} < 6.2$) and Eddington ratios over more than 1. When studying the extreme behavior, the most evident connection is that the low radio luminosity extremes occur on the left (lower) side of Eddington ratio of the base sample, whereas the high radio luminosity extremes occur on the high Eddington ratio (right) side of the base sample. No other clear connection can be seen.

For the respective FWHM(H$\beta$) and $\log(\frac{M_{BH}}{M_{\sun}})$ plot, a positive correlation can be seen and is shown by the correlation coefficient being $r_S \approx 0.61$. The p-value is again 0 and thus the null hypothesis is rejected. For the redshift, there is no clear relation when comparing to FWHM(H$\beta$) as evidenced by $r_S \approx 0.04$.. There is, nonetheless, a clear connection between the sources with extreme behavior in the FWHM(H$\beta$) vs redshift comparison. Sources with extreme radio luminosities are located either below a redshift of 0.1 or above a redshift of $\sim0.4$. The sources at lower redshifts with extreme radio luminosity are sources with very low radio luminosities and subsequently, sources at higher redshifts have very high radio luminosities. Another connection can be found in the extreme variability sources, as they are more common for redshifts above $\sim0.2$, with $\sim79\%$ of the sources with variability having a redshift higher than 0.2. The extendedness extreme is also more likely for sources with FWHM(H$\beta$)>1500 km s$^{-1}$ and z<0.2 with $\sim79\%$ of the extended sources satisfying the criteria.

\textit{F(H$\beta$)} The F(H$\beta$) distribution does not show any clear correlation with Eddington ratio. The correlation rank is $\sim-0.01$ and the p-value is $\sim0.6$, meaning that the null hypothesis cannot be rejected. Sources with extreme low radio luminosities have on average higher F(H$\beta$), with $\sim76\%$ of low extreme radio luminosity sources having a F(H$\beta$) > 500. The high extreme radio luminosities have the opposite relation, as $\sim77\%$ of the high extreme radio luminosity sources have F(H$\beta$) < 500. When comparing the F(H$\beta$) to the logarithmic black hole mass, there is no strong correlation as the correlation rank is $\sim0.2$. For the relation of F(H$\beta$) with the redshift, the correlation rank is $\sim-0.3$, which means that there is moderate negative correlation. The p-value is $\sim0.00$, and thus the null hypothesis is rejected.

\textit{F[O III]} The F[O III] distribution when compared with the Eddington ratio, has a weak negative correlation: $r_S \approx-0.2$. Much like with F(H$\beta$), the stronger F[O III] values, $>500$ are correlated with low extreme radio luminosities, and the weaker F[O III] values, $<500$, are correlated with high extreme radio luminosities. When comparing F[O III] to the logarithmic black hole mass, a very weak negative correlation can be noted: $r_S \approx -0.1$. Finally, when comparing the redshift to F[O III], there is a strong negative correlation as $r_S \approx -0.6$.

\textit{R4570.} We have R4570 results up to a redshift of $\sim0.36$. \citet{2017rakshit}, has given sources with a higher redshift than the aforementioned a result of 0, and these sources will thus not be included in this analysis. This causes several of the extreme sources to not be visible in the plots. When studying the R4570 results, seen in Fig.~\ref{r4570}, the first thing of note in the Eddington ratio plot is that the high radio luminosity extreme sources are virtually all missing. Regarding the correlations, when comparing the Eddington ratio to R4570, a weak correlation can be noted as $r_S \approx 0.2$. When comparing to the black hole mass, there is a very weak negative correlation $r_S \approx-0.1$. No correlation can be noted when comparing R4570 redshift: $r_S \approx-0.01$. Generally a positive correlation should be observable between the R4570 and the redshift; however, this is not visible in our data. Although the trend is clearer towards higher redshifts than the limiting $\sim0.36$, some trending behavior should already be visible \citep{2014shen}. 

\textit{L5100.} A strong positive correlation can be noted when studying Eddington ratio, black hole mass, and redshift with L5100 as the correlation ranks range from $\sim0.5$ to $\sim0.6$. This is as expected as all the properties are related to each other. One clear outlier in the L5100 v. Eddington ratio graph is SDSS J095820.94+322402.2. This source has the highest L5100 luminosity of the sample as well as a very high Eddington ratio ($\sim8.8$). This, however, is most likely due to the fact that the source is jetted \citep[e.g.,][]{2022chen} and thus the L5100 has most likely been overestimated significantly.

\textit{Radio luminosities and sources with extreme behavior.}
If we study how the radio luminosity compares to the L5100, we can see that there is significant scatter for a given AGN luminosities. As an example, for an AGN luminosity of $10^{44}$ erg s$^{-1}$, the radio luminosity varies in the VLASS Epoch 2 plot, seen in Fig.~\ref{rllum1l5100}, from $\sim10^{39}$ erg s$^{-1}$ to $\sim10^{42.8}$ erg s$^{-1}$. This scattering is visible in all the different surveys and is similar to what is observed for more typical Seyferts and quasars. This behavior could provide insight on the question on the origin of the radio emission of the NLS1 galaxies: is the emission caused by star formation or the AGN? It has been suggested that these two emission types can co-exist simultaneously \citep{2019gurkan}. However, no strong conclusions should be drawn due to both incomplete detections as well as the Malmquist bias effect. 

We have also studied how the Eddington ratio and black hole mass compares to the various radio luminosities. Looking at the Eddington ratio plots, seen in Fig.~\ref{rllum1} and Fig.~\ref{rllum2}, a key feature of both the VLASS plots is that the majority of the sources in these plots have at least some type of extreme behavior. There is a larger fraction of non-extreme sources in the FIRST plot compared to either of the VLASS plots. The LoTSS plots have the largest number of non-extreme sources, this is as expected as these are also the largest datasets. Based on these plots, it appears that sources at radio luminosities below $10^{40}$ erg s$^{-1}$ are more likely to only exhibit one type of extreme behavior, whereas sources with higher tend to have at least two types of extreme behavior when such behavior is present. Furthermore, sources with an extreme spectral index flag with an inverted spectrum appear to be more common when $\log{\nu L_{\nu}}$>$\sim$40.9 erg s$^{-1}$ when studying all radio luminosities of these sources. The black hole mass comparisons to the different radio luminosities can be seen in Fig.~\ref{rllumbh1} and Fig.~\ref{rllummbh2}. Generally the extreme behavior does not show any evident correlation with the black hole mass. 

The redshift-radio luminosity plots can be seen in Fig.~\ref{rllumz1} and Fig.~\ref{rllumz2}. If we study the LoTSS vs redshift plot, we can see that on average, excluding the low-end extreme radio luminosity sources, the sources with extreme behavior tend to have the highest radio luminosities in their respective redshift ranges. Thus causing a visible deviation between extreme and non-extreme sources. This can partially be seen also for FIRST, however, this relationship is not visible for either VLASS epoch as the vast majority of sources are extreme.

\section{Discussion}
\label{sec:discussion}

The properties of NLS1 galaxies are rather puzzling. Originally these sources were categorized as radio-quiet; however, this and many other studies have shown this assessment to be faulty \citep[e.g.,][]{2017lahteenmaki1,2018lahteenmaki1}. From a historical point of view the detection fraction of NLS1 galaxies, based on the old samples, has greatly varied. However, as the prior samples have been contaminated, caution should be used when interpreting the results. We are using the historical detection rate as a reference point for this study while keeping in mind the contamination issues. In this section, we will go through the various results from the points of view of general class diversity, origin of the radio emission, variability, as well as extendedness.

\subsection{Class diversity}

Although the NLS1 galaxy population is quite monolithic from an optical spectral perspective when considering AGN in general. When studying the optical plane of eigenvector 1 of NLS1 galaxies, the population is very uniform and located at one extreme of the eigenvector 1 parameter space \citep{2018marziani}. However, this picture breaks down in radio. The radio luminosities range from $\sim10^{35}$ erg s$^{-1}$ to $\sim10^{43}$ erg s$^{-1}$, highlighting the variety of the class, a class that based on its traditional classification system should not present strong radio emission. 

The detection rate we obtained for this sample is clearly different from the historical estimates ($\sim7-30\%$ \citep{2006komossa1,2006zhou}) from FIRST. For FIRST the detection rate of this sample is closer to the lower side of the historical estimate, for which there are various possible causes. One possible explanation is that our sample just happens to include less sources with clear radio emission at the time of observation due to, for example, variability when comparing to historical samples with a higher detection rate. We do note that it is possible that due to the stochastic nature of variability that the variability behavior is seen in prior samples, but due to the same stochastic nature this is unlikely to explain everything. A second possible explanation is those previous samples with a higher detection rate have been contaminated, for example, by BLS1 galaxies that in general have higher radio detection rate than NLS1 galaxies. A complete sample is needed before the accurate detection rate of NLS1 galaxies can be confirmed.

The VLASS detection rates are only $\sim5\%$ and are thus also lower than the detection rate of FIRST. This is not an unexpected result. One possible explanation for this can be in using the Quick Look data. In case a source is barely detected at FIRST (1~mJy) and assuming optically thin emission, then this type of source might not be detectable at 3~GHz as the detection threshold of the VLASS Quick Look data is $\sim600$~$\mu$Jy. Another reason can be that although a source is detected by FIRST, due to the higher resolution of VLASS the diffuse radio emission may be undetectable at 3~GHz. Furthermore the flux density underestimations, which are not taken into account in this study, of the Quick Look data can also affect the detection rate, and finally, variability may also be a part reason for the difference in the detection rates between VLASS and FIRST.

The LoTSS detection rate is the highest of our sample, with almost half of the sample that was within the survey coverage being detected. This cannot be compared to the detection rate of FIRST due to the difference in both frequency and sensitivity. Whereas the LoTSS and VLASS surveys can detect sources at sub-milliJansky flux densities, the sensitivity of FIRST allows only for milliJansky detections. The detection rate of LoTSS is very high with a detection rate of over $44\%$ when taking into account the sky coverage of the survey.

As evidenced by the radio luminosity plots, a considerable number of our sources present high radio luminosities. For FIRST, over $40\%$ of the sample has a radio luminosity above $10^{40}$ erg s$^{-1}$ with $\sim12\%$ having $10^{41}$ erg s$^{-1}$. For VLASS roughly $50\%$ of the sample has a radio luminosity above $10^{40}$ erg s$^{-1}$ and $\sim20\%$ have a radio luminosity above $10^{41}$ erg s$^{-1}$. For LoTSS these statistics are, as expected, lower with only $\sim3-4\%$ of our targets having a radio luminosity above $10^{40}$ erg s$^{-1}$. 

The sources with the weakest radio luminosities are detected by LoTSS. The vast majority of sources detected by LoTSS, and thus majority of the sources in our sample, present radio luminosities below the threshold for starburst and star forming galaxies. Of the sources detected by LoTSS, $\sim90\%$ have a 144~MHz radio luminosity below $10^{39.7}$ erg s$^{-1}$. The strongest radio luminosity of the sample is found for SDSS J095820.94+322402.2, with a radio luminosity of $~10^{43}$ erg s$^{-1}$. Such high radio luminosities are not in agreement with the current NLS1 galaxy classification system. This source is also a $\gamma$-NLS1 \citep{2022chen} and thus most likely hosts powerful relativistic jets.

If we take a look at the spectral properties, the majority of the known relations between the various parameters hold true. The outlier can be found in the redshift-R4570 relation, as no clear relation can be seen although a positive correlation is expected. For the most part, the sample behaves very uniformly from the optical perspective, which is as expected, as the NLS1 galaxy class, as mentioned, is unified from the optical perspective. However, when we study the sources with extreme behavior, we can start spotting some diversity, as discussed below. We do note, that all the spectral results, and the properties reliant on the spectral results, should still be confirmed with higher quality optical data before any final conclusions are to be drawn.

There are a total of 173 sources, $\sim11\%$ of our detected sample, that present some type of extreme behavior, as described in Sect.~\ref{optspec}. Most commonly, for 131 sources, only one type of extreme behavior is present at a time. No source presents all types of different extreme behavior, however, ten sources present all but one type of extreme behavior. One of the most interesting ones of these is SDSS J094857.31+002225.5 as it has been detected at $\gamma$-rays \citep{2009abdo2, 2009abdo1,2009abdo3}. 

From the Eddington ratio perspective, rather curiously the vast majority of the highest Eddington ratio sources do not present any type of extreme behavior. The high Eddington results are, however, a bit suspicious in general, as they may be induced by, for example, jets or possible difficulties in the modeling of the optical spectra. Several earlier studies have claimed that a high Eddington ratio tended to mean non-jettedness, as blazars, for example, have on average an Eddington ratio $< 0.1$ \citep{2014heckman, 2022belladitta}. Nevertheless, this assessment is caused by a selection bias; although the majority of radio AGN have low Eddington ratios, there are plenty of such sources with high Eddington ratios \citep[e.g.,][]{2020yang}. Further proof for this has been found, for example, through simulations using general relativity (radiative) magnetohydrodynamics, where it has been shown that powerful jets can be formed in high Eddington ratio systems when magnetically-arrested accretion is reached \citep{2017mckinney, 2022liska}. Due to this, more extreme behavior could have been expected. We note the lack of extreme sources at high Eddington ratios. However, this does not conclusively indicate that such sources do not exist.

When taking a look at the possible relations of redshift with the different properties, it is important to remember the Malmquist bias as it directly affects some of the relations, both extreme and non-extreme. The Malmquist bias is clearly visible when studying the extreme radio luminosity sources. Based on our plots, sources with a maximum redshift of $\sim0.10$ have the low end extreme radio luminosities, whereas the sources with a redshift of $\sim0.4$ and more have the high end extreme luminosities. The redshift plots highlight the fact, that we are unable to detect high-redshift, low-luminosity sources due to, for example, the sensitivity of our instruments. A similar issue can be noted with the black hole mass, which is expected due to the interplay between the black hole mass, FWHM(H$\beta$), L5100, which in their turn are reliant on the bolometric luminosity, which is again dependent on the redshift.

\subsection{Origin of the radio emission}
\label{radioemission}

For NLS1 galaxies, the origin of the radio emission is behind a curtain of uncertainty. By studying a variety of parameters, such as radio luminosity, spectral index, and variability, we can hopefully start to take a glimpse of what is beyond the curtain \citep{2022jarvela}. From a general perspective, the majority of our sources present a steep spectrum which is in agreement with findings from previous papers \citep[e.g.,][]{2018berton}. The second most common type of spectrum is a flat spectrum. A small number of sources have an inverted spectrum. For us, the most interesting cases include the very steep and the very inverted sources, as these quite often indicate variability induced by a jet. 

In the absence of variability, inverted spectra are generally caused by absorption, either SSA or FFA. However, as we have non-synchronous data, the inverted spectra can also be caused by time variability. One of the highest spectral indices in our sample is for SDSS J110546.06+145202.4 with a spectral index of $\sim4.16$ in the VLASS Epoch 1 v FIRST integrated pair. A similarly high spectral index was found in all the VLASS v FIRST pairs. There are other sources with very inverted spectra: all of these sources can be found in the complete data table.

The highest radio luminosities of our sample, $\log{\nu L_{\nu}} \geq42$ erg s$^{-1}$, are observed in sources detected in VLASS and FIRST. The majority of sources ($\sim51-52\%$) in VLASS have an integrated radio luminosity above the maximum expected star formation luminosity, indicating that in most sources the AGN contributes to the radio emission. In FIRST, the corresponding fraction is $\sim44\%$. The results of LoTSS are all in general relatively low with the overwhelming majority of sources having a radio luminosity below the frequency-dependent thresholds for starburst and star-forming galaxies. It is possible that self-absorption affects some of the AGN emission in the LoTSS band.

Star formation and AGN can both simultaneously affect the radio luminosity of a source \citep{2019gurkan}; however, in the absence of unusual spectral behavior or high-resolution observations it is hard to separate these two different origins. It is possible to give an estimate on the expected radio luminosity of the star formation rate (SFR) through the use of the $M_{*}$-SFR relation \citep{2007elbaz}. This, however, becomes slightly problematic for NLS1 galaxies due to the enhanced star formation \citep{2010sani1} and these sources possibly residing above the main sequence \citep{2010sani1,2015caccianiga,2023salome}. Furthermore, the stellar mass of NLS1 galaxies is known for only a handful of objects. Without these, it is currently impossible to guarantee that the relation used by \citet{2007elbaz} will also hold true for NLS1 galaxies. To get a better estimate, it is necessary to both get direct measurements of the SFR through for example studying the CO emission \citep{2023salome}, and also better estimates of $M_{*}$ of a large sample of NLS1 galaxies. Notwithstanding, with the relation by \citet{2007elbaz}, in the case of NLS1 galaxies being on the main sequence, assuming $M_{BH} = 10^7 M_\sun$ and $M_{*} = 4000 M_{BH}$ \citep{2015reines}, the SFR would be $\sim4.3 M_\sun$ year$^{-1}$. From this, the expected radio luminosity would be $\log{\nu L_{\nu}} \sim37.9$ erg s$^{-1}$, very comparable to what is actually observed for NLS1 with black hole masses at this level. This would suggest that star formation makes up all the radio emission at and below this limit. Such radio luminosities are obtained mostly in the LoTSS data, meaning that some of the LoTSS detected NLS1 galaxies show only star formation with virtually no significant AGN contribution.

The relationship between the radio luminosity and the AGN jet power is strongly dependent on environment and on the type of object as well as on time \citep{2014hardcastle,2018hardcastle} so it is not possible to predict a radio luminosity for a given AGN type. A high AGN and radio luminosity would suggest the presence of relativistic jets; however, some of these results may be caused by Malmquist bias. 

For the spectral index, overall, it is very difficult to state anything on sources that have a spectral index of $\sim-0.8$, as it could be due to optically thin synchrotron either due to star formation or to a jet or outflow, or a combination of these. Generally, we can draw the strongest conclusions about the sources with either very steep or very inverted spectral indices. However, we have several spectral index pairs for which a spectral index in the flat region is the highest spectral index. Due to this we have also studied these sources with extra scrutiny. 

Inverted sources occur more commonly for the VLASS and FIRST pairs than the LoTSS and FIRST pairs. The inverted sources in the VLASS and FIRST pairs may be a result of the long time between the two surveys. For these sources, there is no evident connection with any of the studied parameters. Around $20\%$ of the inverted-spectrum sources also present a variability index above 0.15 for peak flux densities. 

There is no clear correlation between the 33 extremely steep-spectrum sources and any of the studied extreme parameters as these sources are quite well distributed over the various properties. Many of the extremely steep-spectrum sources do, however, present high radio luminosities, suggesting the possibility of compact steep-spectrum sources (CSS) or peaked sources (PS) in general. For FIRST the traditional CSS luminosity threshold is $\log(\nu  L_{\nu}/(erg\ s^{-1})) \geq 41.15$ \citep{1998odea}. When extrapolating to other frequencies, assuming a spectral index of $-1$, the LoTSS threshold is $\log(\nu L_{\nu}/(erg\ s^{-1})) \geq 42.13$, and the VLASS threshold is $\log(\nu  L_{\nu}/(erg\ s^{-1})) \geq 40.82$. There are eight sources that both have extremely steep spectra in either VLASS Epoch 1 v. FIRST (peak) pair or VLASS Epoch 2 v. FIRST (peak) pair, and meet the FIRST and VLASS CSS luminosity thresholds. According to the current paradigm, peaked sources (PS) and CSS sources are typically found in bright, elliptical galaxies with old stellar populations \citep{2021odea}, whereas NLS1 galaxies are predominantly located in disk-like galaxies with pseudo-bulges and have low S\'{e}rsic indices, which generally correlates with low stellar masses \citep{2020sanchez}. However, in a recent study \citet{2024vietri} found that the hosts of low-luminosity CSS sources are disk-like, and thus this population could be related to NLS1 galaxies. Based on an upcoming study, the stellar masses of NLS1 galaxies have been found to be between $10^8$ and $10^{11} M_\sun$, supporting the analysis of lower stellar masses (Syrjärinne et al. in prep.).

\subsection{Variability}

Variability seen in NLS1 galaxies is observationally puzzling due to its irregular nature. These sources have exhibited very short-timescale, but strong variability in radio as well as other frequencies \citep{2024jarvela}. Several sources originally classified as radio-silent have been detected in radio at 37~GHz at surprisingly high flux density levels \citep{2017lahteenmaki1,2018lahteenmaki1}. When studying variability between the two VLASS samples, we note that the different underestimations, as mentioned in Sect.~\ref{sec:sample}, of the flux densities were not taken into account. Due to this, it is possible that, for example, very minor variability is simply caused by the different underestimations. 

Of the 171 sources for which a fractional variability index was determined for 20 sources present a variability index above 0.1 for the integrated flux density results and 16 for the peak flux density results. Furthermore, there are eight sources that present a fractional variability index of above 0.2 for the integrated flux density results and six for the peak flux density results. Sources with this high of a fractional variability index are with the highest odds truly variable. We cannot reliably estimate the number of sources with variability behavior in the other catalogs. 

Another interesting aspect is the study of the sources detected only in one of the VLASS epochs. There are 40 sources detected in VLASS Epoch 1 that are not detected in Epoch 2, whereas there are 36 sources detected in Epoch 2 and not in Epoch 1. All of these sources have relatively low flux densities, ranging from $\sim0.65$~mJy to $\sim6.08$~mJy. As mentioned in Sect.~\ref{sect:fracvar}, when determining the upper limits for the flux density results and then the lower limits for the fractional variability index, many of these sources, 35 for the integrated flux density results and nine for the peak flux density results, could be highly variable (variability index above 0.2). There is a variety of possible reasons for this. The sources could be new radio sources, it could be variability, or it could just be unfortunate timing or various error sources leading to non-detections. In the case of a fractional variability index above 0.2 or for a source that has been just below the detection threshold, it is possible that the lack of detection is a result of observational effects.

\subsection{Extreme extendedness}
When studying extendedness of the VLASS results, there is a large fraction of sources that are clearly point-like. We have considered a source clearly point-like when $S_{\texttt{peak}}/S_{\texttt{integrated}} > 0.95$ \citep{2008kimball}. For VLASS Epoch 1 results $\sim33\%$ of sources satisfy this criteria. The respective fraction of such sources in VLASS Epoch 2 is $\sim25\%$. If a source presents extreme extendedness, it is often the only type of extreme behavior it shows. Out of the 43 extremely extended sources, roughly $\sim37\%$, also show other types of extreme behavior used in this paper (spectral index, radio luminosity, and variability). When studying the deconvolved component major axis size (FWHM) of the sources that have $S_{\texttt{peak}}/ S_{\texttt{integrated}} \leq 0.5$ for the VLASS epochs there are 28 sources that qualify. For Epoch 1, the average size is $\sim3.1"$ and the maximum size is $\sim5.5"$. For Epoch 2 the respective numbers are $\sim3.9"$ and $\sim7.9"$. The most extended source in Epoch 2 was not included in Epoch 1 catalog. The average size of all sources detected with Epoch 1 is $\sim1.3"$ and for Epoch 2 $\sim1.7"$. This means that on average, the most extended sources for both VLASS epoch are more than twice the size of the average source. There is a minor redshift preference toward lower redshifts for extreme extendedness. The mean redshift of sources with extreme extendedness is $\sim0.20$, with only $\sim19\%$ of the sources with extreme extendedness having a redshift over 0.3. This outcome is as expected, as our sources are quite physically small and with increasing redshift, we expected our sources to better fit within the beam of each survey. 

For VLASS, extendedness is more common for sources with lower flux densities. When only accounting for sources with $S_{\texttt{peak}}/S_{\texttt{integrated}} \leq 0.9$, of the 108 sources in VLASS Epoch 1, 58 sources have an integrated flux density of below 2~mJy. For peak flux density, the respective number of sources is 74. For VLASS Epoch 2, the numbers are 65 sources for the integrated flux density and 84 for the peak flux density for a total of 124 sources. For FIRST, this pattern holds true only for the peak flux densities as of the 71 sources, 49 sources have both $S_{\texttt{peak}}/S_{\texttt{integrated}} \leq 0.9$ and a flux density of less than 2~mJy. For the integrated flux densities, 31 sources so $\sim44\%$, qualify.

\section{Conclusions}
\label{sec:conclusions}

We have studied the radio and optical spectral properties of NLS1 galaxies using the cleanest large sample currently available. We used a variety of surveys to obtain as large a coverage as possible and obtained a radio detection for $\sim40\%$ of the sample. The surveys used were: FIRST, NVSS, LoTSS DR 3, and VLASS (Epoch 1 and Epoch 2). Optical spectral parameters were obtained from \citet{2017rakshit}.

\begin{enumerate}
    \item The largest detection rate is obtained in LoTSS and the lowest in VLASS. The strength and contribution of star formation increases towards lower frequencies, with a considerable fraction of LoTSS detections likely being dominated by star formation rather than AGN-related emission.
    \item The radio diversity of the class can be seen in the flux density distribution. The vast majority of our sources present low flux densities at $\ll10$~mJy, as expected, but a fraction have very high flux densities at $>100$~mJy. The high flux densities strongly indicate AGN activity, and contradict the traditional classification system. The optical properties are very unified, as would be expected.
    \item Although the origins of radio emission are often hard to determine, some sources exhibit clear signs of AGN contribution. This can be seen in, for example, significant variability, high radio luminosities, clearly above the star formation limit, and the presence of both very high and very low spectral indices. Extreme behavior is more common towards higher radio luminosities, and for these sources, the AGN activity could be in the form of powerful relativistic jets.
    \item We identified several sources that based on their spectral indices and radio luminosities are CSS candidates. This is further indication of the presence of relativistic jets, and supports the hypothesis that some CSS sources could belong to the parent population of jetted NLS1 galaxies. 
\end{enumerate}

To get a more complete look of these sources, an even cleaner data set is needed. For this, higher resolution optical spectra is a necessity. With higher resolution as well as diligent studying of spectral lines it is possible to weed out the remaining non-NLS1 galaxies such as intermediates and BLS1 galaxies. Based on our results, the historical estimates of the radio properties of the general NLS1 galaxy population is likely to have been overestimated; however, there is a clear group with significant radio emission. With a completely cleaned sample, it will finally be possible to study NLS1 galaxy properties without the interference of other AGN. One area to focus on is the idiosyncratic variability seen in these sources. These sources may provide us a first glance at a new type of variability in both jetted and non-jetted AGN, possibly even a new class of jetted AGN. Furthermore, to better understand where the radio emission comes from in these sources, better estimates of their stellar masses and SFR are needed.

\textit{Data availability}
The full data table is only available in electronic form at the CDS via anonymous ftp to cdsarc.u-strasbg.fr (130.79.128.5) or via http://cdsweb.u-strasbg.fr/cgi-bin/qcat?J/A+A/.

\begin{acknowledgements}

This research has made use of the VizieR catalog access tool, CDS, Strasbourg, France.  The National Radio Astronomy Observatory is a facility of the National Science Foundation operated under cooperative agreement by Associated Universities, Inc. This research has made use of the CIRADA cutout service at URL cutouts.cirada.ca, operated by the Canadian Initiative for Radio Astronomy Data Analysis (CIRADA). CIRADA is funded by a grant from the Canada Foundation for Innovation 2017 Innovation Fund (Project 35999), as well as by the Provinces of Ontario, British Columbia, Alberta, Manitoba and Quebec, in collaboration with the National Research Council of Canada, the US National Radio Astronomy Observatory and Australia’s Commonwealth Scientific and Industrial Research Organisation. The National Radio Astronomy Observatory is a facility of the National Science Foundation operated under cooperative agreement by Associated Universities, Inc. LOFAR data products were provided by the LOFAR Surveys Key Science project (LSKSP; https://lofar-surveys.org/) and were derived from observations with the International LOFAR Telescope (ILT). LOFAR (van Haarlem et al. 2013) is the Low Frequency Array designed and constructed by ASTRON. It has observing, data processing, and data storage facilities in several countries, which are owned by various parties (each with their own funding sources), and which are collectively operated by the ILT foundation under a joint scientific policy. The efforts of the LSKSP have benefited from funding from the European Research Council, NOVA, NWO, CNRS-INSU, the SURF Co-operative, the UK Science and Technology Funding Council and the Jülich Supercomputing Centre. Funding for the Sloan Digital Sky Survey V has been provided by the Alfred P. Sloan Foundation, the Heising-Simons Foundation, the National Science Foundation, and the Participating Institutions. SDSS acknowledges support and resources from the Center for High-Performance Computing at the University of Utah. SDSS telescopes are located at Apache Point Observatory, funded by the Astrophysical Research Consortium and operated by New Mexico State University, and at Las Campanas Observatory, operated by the Carnegie Institution for Science. The SDSS web site is \url{www.sdss.org}. SDSS is managed by the Astrophysical Research Consortium for the Participating Institutions of the SDSS Collaboration, including Caltech, The Carnegie Institution for Science, Chilean National Time Allocation Committee (CNTAC) ratified researchers, The Flatiron Institute, the Gotham Participation Group, Harvard University, Heidelberg University, The Johns Hopkins University, L’Ecole polytechnique f\'{e}d\'{e}rale de Lausanne (EPFL), Leibniz-Institut f\"{u}r Astrophysik Potsdam (AIP), Max-Planck-Institut f\"{u}r Astronomie (MPIA Heidelberg), Max-Planck-Institut f\"{u}r Extraterrestrische Physik (MPE), Nanjing University, National Astronomical Observatories of China (NAOC), New Mexico State University, The Ohio State University, Pennsylvania State University, Smithsonian Astrophysical Observatory, Space Telescope Science Institute (STScI), the Stellar Astrophysics Participation Group, Universidad Nacional Aut\'{o}noma de M\'{e}xico, University of Arizona, University of Colorado Boulder, University of Illinois at Urbana-Champaign, University of Toronto, University of Utah, University of Virginia, Yale University, and Yunnan University. IV wants to thank the Magnus Ehrnrooth Foundation for their continuing support. MJH thanks the UK Science and Technology Facilities Council (STFC) for support [[ST/V000624/1]].

\end{acknowledgements}

\bibliographystyle{aa}
\bibliography{aa50962-24.bib}

\begin{thebibliography}{85}
\expandafter\ifx\csname natexlab\endcsname\relax\def\natexlab#1{#1}\fi

\bibitem[{{Abdo} {et~al.}(2009{\natexlab{a}}){Abdo}, {Ackermann}, {Ajello}, {Axelsson}, {Baldini}, {Ballet}, {Barbiellini}, {Bastieri}, {Battelino}, {Baughman}, {Bechtol}, {Bellazzini}, {Bloom}, {Bonamente}, {Borgland}, {Bregeon}, {Brez}, {Brigida}, {Bruel}, {Caliandro}, {Cameron}, {Caraveo}, {Casandjian}, {Cavazzuti}, {Cecchi}, {Chekhtman}, {Cheung}, {Chiang}, {Ciprini}, {Claus}, {Cohen-Tanugi}, {Collmar}, {Conrad}, {Costamante}, {Dermer}, {de Angelis}, {de Palma}, {Digel}, {Silva}, {Drell}, {Dubois}, {Dumora}, {Farnier}, {Favuzzi}, {Focke}, {Foschini}, {Frailis}, {Fuhrmann}, {Fukazawa}, {Funk}, {Fusco}, {Gargano}, {Gehrels}, {Germani}, {Giebels}, {Giglietto}, {Giordano}, {Giroletti}, {Glanzman}, {Grenier}, {Grondin}, {Grove}, {Guillemot}, {Guiriec}, {Hanabata}, {Harding}, {Hartman}, {Hayashida}, {Hays}, {Hughes}, {J{\'o}hannesson}, {Johnson}, {Johnson}, {Johnson}, {Kamae}, {Katagiri}, {Kataoka}, {Kerr}, {Kn{\"o}dlseder}, {Kuehn}, {Kuss}, {Lande}, {Latronico}, {Lemoine-Goumard}, {Longo}, {Loparco}, {Lott},
  {Lovellette}, {Lubrano}, {Madejski}, {Makeev}, {Max-Moerbeck}, {Mazziotta}, {McConville}, {McEnery}, {Meurer}, {Michelson}, {Mitthumsiri}, {Mizuno}, {Monte}, {Monzani}, {Morselli}, {Moskalenko}, {Murgia}, {Nolan}, {Norris}, {Nuss}, {Ohsugi}, {Omodei}, {Orlando}, {Ormes}, {Paneque}, {Panetta}, {Parent}, {Pavlidou}, {Pearson}, {Pepe}, {Pesce-Rollins}, {Piron}, {Porter}, {Rain{\`o}}, {Rando}, {Razzano}, {Readhead}, {Reimer}, {Reimer}, {Reposeur}, {Richards}, {Ritz}, {Rodriguez}, {Romani}, {Ryde}, {Sadrozinski}, {Sambruna}, {Sanchez}, {Sander}, {Parkinson}, {Scargle}, {Schalk}, {Sgr{\`o}}, {Smith}, {Spandre}, {Spinelli}, {Starck}, {Stevenson}, {Strickman}, {Suson}, {Tagliaferri}, {Takahashi}, {Tanaka}, {Thayer}, {Thompson}, {Tibaldo}, {Tibolla}, {Torres}, {Tosti}, {Tramacere}, {Uchiyama}, {Usher}, {Vilchez}, {Vitale}, {Waite}, {Winer}, {Wood}, {Ylinen}, {Zensus}, {Ziegler}, {Fermi/LAT Collaboration}, {Ghisellini}, {Maraschi}, {Tavecchio}, \& {Angelakis}}]{2009abdo2}
{Abdo}, A.~A., {Ackermann}, M., {Ajello}, M., {et~al.} 2009{\natexlab{a}}, \apj, 699, 976

\bibitem[{{Abdo} {et~al.}(2009{\natexlab{b}}){Abdo}, {Ackermann}, {Ajello}, {Axelsson}, {Baldini}, {Ballet}, {Barbiellini}, {Bastieri}, {Baughman}, {Bechtol}, \& et~al.}]{2009abdo1}
{Abdo}, A.~A., {Ackermann}, M., {Ajello}, M., {et~al.} 2009{\natexlab{b}}, \apj, 707, 727

\bibitem[{{Abdo} {et~al.}(2009{\natexlab{c}}){Abdo}, {Ackermann}, {Ajello}, {Baldini}, {Ballet}, {Barbiellini}, {Bastieri}, {Bechtol}, {Bellazzini}, {Berenji}, {Bloom}, {Bonamente}, {Borgland}, {Bregeon}, {Brez}, {Brigida}, {Bruel}, {Burnett}, {Caliandro}, {Cameron}, {Caraveo}, {Casandjian}, {Cecchi}, {{\c{C}}elik}, {Chekhtman}, {Cheung}, {Chiang}, {Ciprini}, {Claus}, {Cohen-Tanugi}, {Conrad}, {Cutini}, {Dermer}, {de Palma}, {Silva}, {Drell}, {Dubois}, {Dumora}, {Farnier}, {Favuzzi}, {Fegan}, {Focke}, {Foschini}, {Frailis}, {Fukazawa}, {Fusco}, {Gargano}, {Gehrels}, {Germani}, {Giebels}, {Giglietto}, {Giordano}, {Giroletti}, {Glanzman}, {Godfrey}, {Grenier}, {Grove}, {Guillemot}, {Guiriec}, {Hayashida}, {Hays}, {Horan}, {Hughes}, {J{\'o}hannesson}, {Johnson}, {Johnson}, {Kadler}, {Kamae}, {Katagiri}, {Kataoka}, {Kerr}, {Kn{\"o}dlseder}, {Kuss}, {Lande}, {Latronico}, {Longo}, {Loparco}, {Lott}, {Lovellette}, {Lubrano}, {Makeev}, {Mazziotta}, {McConville}, {McEnery}, {Meurer}, {Michelson}, {Mitthumsiri},
  {Mizuno}, {Monte}, {Monzani}, {Morselli}, {Moskalenko}, {Murgia}, {Nolan}, {Norris}, {Nuss}, {Ohsugi}, {Omodei}, {Orlando}, {Ormes}, {Pelassa}, {Pepe}, {Persic}, {Pesce-Rollins}, {Piron}, {Porter}, {Rain{\`o}}, {Rando}, {Razzano}, {Rochester}, {Rodriguez}, {Ryde}, {Sadrozinski}, {Sambruna}, {Sander}, {Saz Parkinson}, {Scargle}, {Sgr{\`o}}, {Smith}, {Spandre}, {Spinelli}, {Strickman}, {Suson}, {Tagliaferri}, {Takahashi}, {Takahashi}, {Tanaka}, {Thayer}, {Thayer}, {Thompson}, {Tibaldo}, {Tibolla}, {Torres}, {Tosti}, {Tramacere}, {Uchiyama}, {Usher}, {Vasileiou}, {Vilchez}, {Vitale}, {Waite}, {Wang}, {Winer}, {Wood}, {Ylinen}, {Ziegler}, {Fermi/LAT Collaboration}, {Ghisellini}, {Maraschi}, \& {Tavecchio}}]{2009abdo3}
{Abdo}, A.~A., {Ackermann}, M., {Ajello}, M., {et~al.} 2009{\natexlab{c}}, \apjl, 707, L142

\bibitem[{{Aller} {et~al.}(1992){Aller}, {Aller}, \& {Hughes}}]{1992aller}
{Aller}, M.~F., {Aller}, H.~D., \& {Hughes}, P.~A. 1992, \apj, 399, 16

\bibitem[{{Becker} {et~al.}(1994){Becker}, {White}, \& {Helfand}}]{1994becker}
{Becker}, R.~H., {White}, R.~L., \& {Helfand}, D.~J. 1994, in Astronomical Society of the Pacific Conference Series, Vol.~61, Astronomical Data Analysis Software and Systems III, ed. D.~R. {Crabtree}, R.~J. {Hanisch}, \& J.~{Barnes}, 165

\bibitem[{{Belladitta} {et~al.}(2022){Belladitta}, {Caccianiga}, {Diana}, {Moretti}, {Severgnini}, {Pedani}, {Cassar{\`a}}, {Spingola}, {Ighina}, {Rossi}, \& {Della Ceca}}]{2022belladitta}
{Belladitta}, S., {Caccianiga}, A., {Diana}, A., {et~al.} 2022, \aap, 660, A74

\bibitem[{{Bentz} {et~al.}(2013){Bentz}, {Denney}, {Grier}, {Barth}, {Peterson}, {Vestergaard}, {Bennert}, {Canalizo}, {De Rosa}, {Filippenko}, {Gates}, {Greene}, {Li}, {Malkan}, {Pogge}, {Stern}, {Treu}, \& {Woo}}]{2013bentz}
{Bentz}, M.~C., {Denney}, K.~D., {Grier}, C.~J., {et~al.} 2013, \apj, 767, 149

\bibitem[{{Berton} {et~al.}(2020{\natexlab{a}}){Berton}, {Bj{\"o}rklund}, {L{\"a}hteenm{\"a}ki}, {Congiu}, {J{\"a}rvel{\"a}}, {Terreran}, \& {La Mura}}]{2020berton2}
{Berton}, M., {Bj{\"o}rklund}, I., {L{\"a}hteenm{\"a}ki}, A., {et~al.} 2020{\natexlab{a}}, Contributions of the Astronomical Observatory Skalnate Pleso, 50, 270

\bibitem[{{Berton} {et~al.}(2016){Berton}, {Caccianiga}, {Foschini}, {Peterson}, {Mathur}, {Terreran}, {Ciroi}, {Congiu}, {Cracco}, {Frezzato}, {La Mura}, \& {Rafanelli}}]{2016berton}
{Berton}, M., {Caccianiga}, A., {Foschini}, L., {et~al.} 2016, \aap, 591, A98

\bibitem[{{Berton} {et~al.}(2018){Berton}, {Congiu}, {J{\"a}rvel{\"a}}, {Antonucci}, {Kharb}, {Lister}, {Tarchi}, {Caccianiga}, {Chen}, {Foschini}, {L{\"a}hteenm{\"a}ki}, {Richards}, {Ciroi}, {Cracco}, {Frezzato}, {La Mura}, \& {Rafanelli}}]{2018berton}
{Berton}, M., {Congiu}, E., {J{\"a}rvel{\"a}}, E., {et~al.} 2018, \aap, 614, A87

\bibitem[{{Berton} \& {J{\"a}rvel{\"a}}(2021)}]{2021berton2}
{Berton}, M. \& {J{\"a}rvel{\"a}}, E. 2021, Universe, 7, 188

\bibitem[{{Berton} {et~al.}(2020{\natexlab{b}}){Berton}, {J{\"a}rvel{\"a}}, {Crepaldi}, {L{\"a}hteenm{\"a}ki}, {Tornikoski}, {Congiu}, {Kharb}, {Terreran}, \& {Vietri}}]{2020berton}
{Berton}, M., {J{\"a}rvel{\"a}}, E., {Crepaldi}, L., {et~al.} 2020{\natexlab{b}}, \aap, 636, A64

\bibitem[{{Boroson}(2005)}]{2005boroson}
{Boroson}, T. 2005, \aj, 130, 381

\bibitem[{{Boroson} \& {Green}(1992)}]{1992boroson}
{Boroson}, T.~A. \& {Green}, R.~F. 1992, \apj, 80, 109

\bibitem[{{Caccianiga} {et~al.}(2015){Caccianiga}, {Ant{\'o}n}, {Ballo}, {Foschini}, {Maccacaro}, {Della Ceca}, {Severgnini}, {March{\~a}}, {Mateos}, \& {Sani}}]{2015caccianiga}
{Caccianiga}, A., {Ant{\'o}n}, S., {Ballo}, L., {et~al.} 2015, \mnras, 451, 1795

\bibitem[{{Chen} {et~al.}(2018){Chen}, {Berton}, {La Mura}, {Congiu}, {Foschini}, H., {Ciroi}, {Rafanelli}, \& {Bastieri}}]{2018chen}
{Chen}, S., {Berton}, M., {La Mura}, G., {et~al.} 2018, \aap, 615, A167

\bibitem[{{Chen} {et~al.}(2022){Chen}, {Gu}, {Fan}, {Yu}, {Ding}, {Guo}, \& {Xiong}}]{2022chen}
{Chen}, Y., {Gu}, Q., {Fan}, J., {et~al.} 2022, \mnras, 517, 1381

\bibitem[{{Condon} {et~al.}(1998){Condon}, {Cotton}, {Greisen}, {Yin}, {Perley}, {Taylor}, \& {Broderick}}]{1998condon}
{Condon}, J.~J., {Cotton}, W.~D., {Greisen}, E.~W., {et~al.} 1998, \aj, 115, 1693

\bibitem[{{Cracco} {et~al.}(2016){Cracco}, {Ciroi}, {Berton}, {Di Mille}, {Foschini}, {La Mura}, \& {Rafanelli}}]{2016cracco1}
{Cracco}, V., {Ciroi}, S., {Berton}, M., {et~al.} 2016, \mnras, 462, 1256

\bibitem[{{Du} {et~al.}(2018){Du}, {Brotherton}, {Wang}, {Huang}, {Hu}, {Kasper}, {Chick}, {Nguyen}, {Maithil}, {Hand}, {Li}, {Ho}, {Bai}, {Bian}, {Wang}, \& {MAHA Collaboration}}]{2018du}
{Du}, P., {Brotherton}, M.~S., {Wang}, K., {et~al.} 2018, \apj, 869, 142

\bibitem[{{Du} \& {Wang}(2019)}]{2019du1}
{Du}, P. \& {Wang}, J.-M. 2019, \apj, 886, 42

\bibitem[{{Elbaz} {et~al.}(2007){Elbaz}, {Daddi}, {Le Borgne}, {Dickinson}, {Alexander}, {Chary}, {Starck}, {Brandt}, {Kitzbichler}, {MacDonald}, {Nonino}, {Popesso}, {Stern}, \& {Vanzella}}]{2007elbaz}
{Elbaz}, D., {Daddi}, E., {Le Borgne}, D., {et~al.} 2007, \aap, 468, 33

\bibitem[{{Fan}(2020)}]{2020fan}
{Fan}, X.-L. 2020, Universe, 6, 45

\bibitem[{{Foschini}(2011)}]{2011foschini}
{Foschini}, L. 2011, in Narrow-Line Seyfert 1 Galaxies and their Place in the Universe, ed. L.~{Foschini}, M.~{Colpi}, L.~{Gallo}, D.~{Grupe}, S.~{Komossa}, K.~{Leighly}, \& S.~{Mathur}, 24

\bibitem[{{Foschini} {et~al.}(2015){Foschini}, {Berton}, {Caccianiga}, {Ciroi}, {Cracco}, {Peterson}, {Angelakis}, {Braito}, {Fuhrmann}, {Gallo}, {Grupe}, {J{\"a}rvel{\"a}}, {Kaufmann}, {Komossa}, {Kovalev}, {L{\"a}hteenm{\"a}ki}, {Lisakov}, {Lister}, {Mathur}, {Richards}, {Romano}, {Sievers}, {Tagliaferri}, {Tammi}, {Tibolla}, {Tornikoski}, {Vercellone}, {La Mura}, {Maraschi}, \& {Rafanelli}}]{2015foschini1}
{Foschini}, L., {Berton}, M., {Caccianiga}, A., {et~al.} 2015, \aap, 575, A13

\bibitem[{{Foschini} {et~al.}(2010){Foschini}, {Fermi/Lat Collaboration}, {Ghisellini}, {Maraschi}, {Tavecchio}, \& {Angelakis}}]{2010foschini}
{Foschini}, L., {Fermi/Lat Collaboration}, {Ghisellini}, G., {et~al.} 2010, in Astronomical Society of the Pacific Conference Series, Vol. 427, Accretion and Ejection in AGN: a Global View, ed. L.~{Maraschi}, G.~{Ghisellini}, R.~{Della Ceca}, \& F.~{Tavecchio}, 243--248

\bibitem[{{Gliozzi} \& {Williams}(2020)}]{2020gliozzi}
{Gliozzi}, M. \& {Williams}, J.~K. 2020, \mnras, 491, 532

\bibitem[{{Goodrich}(1989)}]{1989goodrich1}
{Goodrich}, R.~W. 1989, \apj, 342, 224

\bibitem[{{Gordon} {et~al.}(2021){Gordon}, {Boyce}, {O'Dea}, {Rudnick}, {Andernach}, {Vantyghem}, {Baum}, {Bui}, {Dionyssiou}, {Safi-Harb}, \& {Sander}}]{2021gordon}
{Gordon}, Y.~A., {Boyce}, M.~M., {O'Dea}, C.~P., {et~al.} 2021, \apjs, 255, 30

\bibitem[{{G{\"u}rkan} {et~al.}(2019){G{\"u}rkan}, {Hardcastle}, {Best}, {Morabito}, {Prandoni}, {Jarvis}, {Duncan}, {Calistro Rivera}, {Callingham}, {Cochrane}, {Croston}, {Heald}, {Mingo}, {Mooney}, {Sabater}, {R{\"o}ttgering}, {Shimwell}, {Smith}, {Tasse}, \& {Williams}}]{2019gurkan}
{G{\"u}rkan}, G., {Hardcastle}, M.~J., {Best}, P.~N., {et~al.} 2019, \aap, 622, A11

\bibitem[{{Hardcastle}(2018)}]{2018hardcastle}
{Hardcastle}, M.~J. 2018, \mnras, 475, 2768

\bibitem[{{Hardcastle} \& {Krause}(2014)}]{2014hardcastle}
{Hardcastle}, M.~J. \& {Krause}, M.~G.~H. 2014, \mnras, 443, 1482

\bibitem[{{Heckman} \& {Best}(2014)}]{2014heckman}
{Heckman}, T.~M. \& {Best}, P.~N. 2014, \araa, 52, 589

\bibitem[{{Helfand} {et~al.}(2015){Helfand}, {White}, \& {Becker}}]{2015helfand}
{Helfand}, D.~J., {White}, R.~L., \& {Becker}, R.~H. 2015, \apj, 801, 26

\bibitem[{{Huang} {et~al.}(2017){Huang}, {Zhang}, {Zhong}, \& {Liang}}]{2017huang}
{Huang}, X.-L., {Zhang}, H.-M., {Zhong}, S.-Q., \& {Liang}, E.-W. 2017, in New Frontiers in Black Hole Astrophysics, ed. A.~{Gomboc}, Vol. 324, 82--84

\bibitem[{{J{\"a}rvel{\"a}} {et~al.}(2020){J{\"a}rvel{\"a}}, {Berton}, {Ciroi}, {Congiu}, {L{\"a}hteenm{\"a}ki}, \& {Di Mille}}]{2020jarvela}
{J{\"a}rvel{\"a}}, E., {Berton}, M., {Ciroi}, S., {et~al.} 2020, \aap, 636, L12

\bibitem[{{J{\"a}rvel{\"a}} {et~al.}(2021){J{\"a}rvel{\"a}}, {Berton}, \& {Crepaldi}}]{2021jarvela}
{J{\"a}rvel{\"a}}, E., {Berton}, M., \& {Crepaldi}, L. 2021, Frontiers in Astronomy and Space Sciences, 8, 147

\bibitem[{{J{\"a}rvel{\"a}} {et~al.}(2022){J{\"a}rvel{\"a}}, {Dahale}, {Crepaldi}, {Berton}, {Congiu}, \& {Antonucci}}]{2022jarvela}
{J{\"a}rvel{\"a}}, E., {Dahale}, R., {Crepaldi}, L., {et~al.} 2022, \aap, 658, A12

\bibitem[{{J{\"a}rvel{\"a}} {et~al.}(2018){J{\"a}rvel{\"a}}, {L{\"a}hteenm{\"a}ki}, \& {Berton}}]{2018jarvela}
{J{\"a}rvel{\"a}}, E., {L{\"a}hteenm{\"a}ki}, A., \& {Berton}, M. 2018, \aap, 619, A69

\bibitem[{{J{\"a}rvel{\"a}} {et~al.}(2017){J{\"a}rvel{\"a}}, {L{\"a}hteenm{\"a}ki}, {Lietzen}, {Poudel}, {Hein{\"a}m{\"a}ki}, \& {Einasto}}]{2017jarvela}
{J{\"a}rvel{\"a}}, E., {L{\"a}hteenm{\"a}ki}, A., {Lietzen}, H., {et~al.} 2017, \aap, 606, A9

\bibitem[{{J{\"a}rvel{\"a}} {et~al.}(2015){J{\"a}rvel{\"a}}, {L{\"a}hteenm{\"a}ki}, \& {Léon-Tavares}}]{2015jarvela}
{J{\"a}rvel{\"a}}, E., {L{\"a}hteenm{\"a}ki}, A., \& {Léon-Tavares}, J. 2015, \aap, 573, A76

\bibitem[{{J{\"a}rvel{\"a}} {et~al.}(2024){J{\"a}rvel{\"a}}, {Savolainen}, {Berton}, {L{\"a}hteenm{\"a}ki}, {Kiehlmann}, {Hovatta}, {Varglund}, {Readhead}, {Tornikoski}, {Max-Moerbeck}, {Reeves}, \& {Suutarinen}}]{2024jarvela}
{J{\"a}rvel{\"a}}, E., {Savolainen}, T., {Berton}, M., {et~al.} 2024, \mnras, 532, 3069

\bibitem[{{Kimball} \& {Ivezi{\'c}}(2008)}]{2008kimball}
{Kimball}, A.~E. \& {Ivezi{\'c}}, {\v{Z}}. 2008, \aj, 136, 684

\bibitem[{{Klimek} {et~al.}(2004){Klimek}, {Gaskell}, \& {Hedrick}}]{2004klimek}
{Klimek}, E.~S., {Gaskell}, C.~M., \& {Hedrick}, C.~H. 2004, \apj, 609, 69

\bibitem[{{Komossa} {et~al.}(2006){Komossa}, {Voges}, {Xu}, {Mathur}, {Adorf}, {Lemson}, {Duschl}, \& {Grupe}}]{2006komossa1}
{Komossa}, S., {Voges}, W., {Xu}, D., {et~al.} 2006, \aj, 132, 531

\bibitem[{{Komossa} {et~al.}(2008){Komossa}, {Xu}, {Zhou}, {Storchi-Bergmann}, \& {Binette}}]{2008komossa}
{Komossa}, S., {Xu}, D., {Zhou}, H., {Storchi-Bergmann}, T., \& {Binette}, L. 2008, \apj, 680, 926

\bibitem[{{Kova{\v{c}}evi{\'c}} {et~al.}(2010){Kova{\v{c}}evi{\'c}}, {Popovi{\'c}}, \& {Dimitrijevi{\'c}}}]{2010kovacevic}
{Kova{\v{c}}evi{\'c}}, J., {Popovi{\'c}}, L.~{\v{C}}., \& {Dimitrijevi{\'c}}, M.~S. 2010, \apjs, 189, 15

\bibitem[{{Kuckartz} {et~al.}(2013){Kuckartz}, {Rädiker}, {Ebert}, \& {Schehl}}]{kuckartz2013}
{Kuckartz}, U., {Rädiker}, S., {Ebert}, T., \& {Schehl}, J. 2013, Statistik: Eine verstaandliche Einfuuhrung (VS Verlag fuur Sozialwissenschaften)

\bibitem[{{Lacy} {et~al.}(2020){Lacy}, {Baum}, {Chandler}, {Chatterjee}, {Clarke}, {Deustua}, {English}, {Farnes}, {Gaensler}, {Gugliucci}, {Hallinan}, {Kent}, {Kimball}, {Law}, {Lazio}, {Marvil}, {Mao}, {Medlin}, {Mooley}, {Murphy}, {Myers}, {Osten}, {Richards}, {Rosolowsky}, {Rudnick}, {Schinzel}, {Sivakoff}, {Sjouwerman}, {Taylor}, {White}, {Wrobel}, {Andernach}, {Beasley}, {Berger}, {Bhatnager}, {Birkinshaw}, {Bower}, {Brandt}, {Brown}, {Burke-Spolaor}, {Butler}, {Comerford}, {Demorest}, {Fu}, {Giacintucci}, {Golap}, {G{\"u}th}, {Hales}, {Hiriart}, {Hodge}, {Horesh}, {Ivezi{\'c}}, {Jarvis}, {Kamble}, {Kassim}, {Liu}, {Loinard}, {Lyons}, {Masters}, {Mezcua}, {Moellenbrock}, {Mroczkowski}, {Nyland}, {O'Dea}, {O'Sullivan}, {Peters}, {Radford}, {Rao}, {Robnett}, {Salcido}, {Shen}, {Sobotka}, {Witz}, {Vaccari}, {van Weeren}, {Vargas}, {Williams}, \& {Yoon}}]{2020lacy}
{Lacy}, M., {Baum}, S.~A., {Chandler}, C.~J., {et~al.} 2020, \pasp, 132, 035001

\bibitem[{{L{\"a}hteenm{\"a}ki} {et~al.}(2017){L{\"a}hteenm{\"a}ki}, {J{\"a}rvel{\"a}}, {Hovatta}, {Tornikoski}, {Harrison}, {L{\'o}pez- Caniego}, {Max-Moerbeck}, {Mingaliev}, {Pearson}, {Ramakrishnan}, {Readhead}, {Reeves}, {Richards}, {Sotnikova}, \& {Tammi}}]{2017lahteenmaki1}
{L{\"a}hteenm{\"a}ki}, A., {J{\"a}rvel{\"a}}, E., {Hovatta}, T., {et~al.} 2017, \aap, 603, A100

\bibitem[{{L{\"a}hteenm{\"a}ki} {et~al.}(2018){L{\"a}hteenm{\"a}ki}, {J{\"a}rvel{\"a}}, {Ramakrishnan}, {Tornikoski}, {Tammi}, {Vera}, \& {Chamani}}]{2018lahteenmaki1}
{L{\"a}hteenm{\"a}ki}, A., {J{\"a}rvel{\"a}}, E., {Ramakrishnan}, V., {et~al.} 2018, \aap, 614, L1

\bibitem[{{Laor}(2000)}]{2000laor1}
{Laor}, A. 2000, \apjl, 543, L111

\bibitem[{{Liska} {et~al.}(2022){Liska}, {Musoke}, {Tchekhovskoy}, {Porth}, \& {Beloborodov}}]{2022liska}
{Liska}, M.~T.~P., {Musoke}, G., {Tchekhovskoy}, A., {Porth}, O., \& {Beloborodov}, A.~M. 2022, \apjl, 935, L1

\bibitem[{{Marziani} {et~al.}(2018){Marziani}, {Dultzin}, {Sulentic}, {Del Olmo}, {Negrete}, {Mart{\'\i}nez-Aldama}, {D'Onofrio}, {Bon}, {Bon}, \& {Stirpe}}]{2018marziani}
{Marziani}, P., {Dultzin}, D., {Sulentic}, J.~W., {et~al.} 2018, Frontiers in Astronomy and Space Sciences, 5, 6

\bibitem[{{Mathur}(2000)}]{2000mathur}
{Mathur}, S. 2000, \mnras, 314, L17

\bibitem[{{McKinney} {et~al.}(2017){McKinney}, {Chluba}, {Wielgus}, {Narayan}, \& {Sadowski}}]{2017mckinney}
{McKinney}, J.~C., {Chluba}, J., {Wielgus}, M., {Narayan}, R., \& {Sadowski}, A. 2017, \mnras, 467, 2241

\bibitem[{{O'Dea}(1998)}]{1998odea}
{O'Dea}, C.~P. 1998, \pasp, 110, 493

\bibitem[{{O'Dea} \& {Saikia}(2021)}]{2021odea}
{O'Dea}, C.~P. \& {Saikia}, D.~J. 2021, \aapr, 29, 3

\bibitem[{{Orban de Xivry} {et~al.}(2011){Orban de Xivry}, {Davies}, {Schartmann}, {Komossa}, {Marconi}, {Hicks}, {Engel}, \& {Tacconi}}]{2011orban}
{Orban de Xivry}, G., {Davies}, R., {Schartmann}, M., {et~al.} 2011, \mnras, 417, 2721

\bibitem[{{Osterbrock} \& {Pogge}(1985)}]{1985osterbrock1}
{Osterbrock}, D.~E. \& {Pogge}, R.~W. 1985, \apj, 297, 166

\bibitem[{{Paliya} {et~al.}(2024){Paliya}, {Stalin}, {Dom{\'\i}nguez}, \& {Saikia}}]{2024paliya}
{Paliya}, V.~S., {Stalin}, C.~S., {Dom{\'\i}nguez}, A., \& {Saikia}, D.~J. 2024, \mnras, 527, 7055

\bibitem[{{Peterson}(2011)}]{2011peterson}
{Peterson}, B.~M. 2011, ArXiv:1109.4181 [\eprint[arXiv]{1109.4181}]

\bibitem[{{Rakshit} {et~al.}(2017){Rakshit}, {Stalin}, {Chand}, \& {Zhang}}]{2017rakshit}
{Rakshit}, S., {Stalin}, C.~S., {Chand}, H., \& {Zhang}, X.-G. 2017, \apjs, 229, 39

\bibitem[{{Reines} \& {Volonteri}(2015)}]{2015reines}
{Reines}, A.~E. \& {Volonteri}, M. 2015, \apj, 813, 82

\bibitem[{{Salom{\'e}} {et~al.}(2023){Salom{\'e}}, {Krongold}, {Longinotti}, {Bischetti}, {Garc{\'\i}a-Burillo}, {Vega}, {S{\'a}nchez-Portal}, {Feruglio}, {Jim{\'e}nez-Donaire}, \& {Zanchettin}}]{2023salome}
{Salom{\'e}}, Q., {Krongold}, Y., {Longinotti}, A.~L., {et~al.} 2023, \mnras, 524, 3130

\bibitem[{{S{\'a}nchez Almeida}(2020)}]{2020sanchez}
{S{\'a}nchez Almeida}, J. 2020, \mnras, 495, 78

\bibitem[{{Sani} {et~al.}(2010){Sani}, {Lutz}, {Risaliti}, {Netzer}, {Gallo}, {Trakhtenbrot}, {Sturm}, \& {Boller}}]{2010sani1}
{Sani}, E., {Lutz}, D., {Risaliti}, G., {et~al.} 2010, \mnras, 403, 1246

\bibitem[{{Sargsyan} \& {Weedman}(2009)}]{2009sargsyan}
{Sargsyan}, L.~A. \& {Weedman}, D.~W. 2009, \apj, 701, 1398

\bibitem[{{Shen} \& {Ho}(2014)}]{2014shen}
{Shen}, Y. \& {Ho}, L.~C. 2014, \nat, 513, 210

\bibitem[{{Shimwell} {et~al.}(2022){Shimwell}, {Hardcastle}, {Tasse}, {Best}, {R{\"o}ttgering}, {Williams}, {Botteon}, {Drabent}, {Mechev}, {Shulevski}, {van Weeren}, {Bester}, {Br{\"u}ggen}, {Brunetti}, {Callingham}, {Chy{\.z}y}, {Conway}, {Dijkema}, {Duncan}, {de Gasperin}, {Hale}, {Haverkorn}, {Hugo}, {Jackson}, {Mevius}, {Miley}, {Morabito}, {Morganti}, {Offringa}, {Oonk}, {Rafferty}, {Sabater}, {Smith}, {Schwarz}, {Smirnov}, {O'Sullivan}, {Vedantham}, {White}, {Albert}, {Alegre}, {Asabere}, {Bacon}, {Bonafede}, {Bonnassieux}, {Brienza}, {Bilicki}, {Bonato}, {Calistro Rivera}, {Cassano}, {Cochrane}, {Croston}, {Cuciti}, {Dallacasa}, {Danezi}, {Dettmar}, {Di Gennaro}, {Edler}, {En{\ss}lin}, {Emig}, {Franzen}, {Garc{\'\i}a-Vergara}, {Grange}, {G{\"u}rkan}, {Hajduk}, {Heald}, {Heesen}, {Hoang}, {Hoeft}, {Horellou}, {Iacobelli}, {Jamrozy}, {Jeli{\'c}}, {Kondapally}, {Kukreti}, {Kunert-Bajraszewska}, {Magliocchetti}, {Mahatma}, {Ma{\l}ek}, {Mandal}, {Massaro}, {Meyer-Zhao}, {Mingo}, {Mostert}, {Nair},
  {Nakoneczny}, {Nikiel-Wroczy{\'n}ski}, {Orr{\'u}}, {Pajdosz-{\'S}mierciak}, {Pasini}, {Prandoni}, {van Piggelen}, {Rajpurohit}, {Retana-Montenegro}, {Riseley}, {Rowlinson}, {Saxena}, {Schrijvers}, {Sweijen}, {Siewert}, {Timmerman}, {Vaccari}, {Vink}, {West}, {Wo{\l}owska}, {Zhang}, \& {Zheng}}]{2022shimwell}
{Shimwell}, T.~W., {Hardcastle}, M.~J., {Tasse}, C., {et~al.} 2022, \aap, 659, A1

\bibitem[{{Shimwell} {et~al.}(2017){Shimwell}, {R{\"o}ttgering}, {Best}, {Williams}, {Dijkema}, {de Gasperin}, {Hardcastle}, {Heald}, {Hoang}, {Horneffer}, {Intema}, {Mahony}, {Mandal}, {Mechev}, {Morabito}, {Oonk}, {Rafferty}, {Retana-Montenegro}, {Sabater}, {Tasse}, {van Weeren}, {Br{\"u}ggen}, {Brunetti}, {Chy{\.z}y}, {Conway}, {Haverkorn}, {Jackson}, {Jarvis}, {McKean}, {Miley}, {Morganti}, {White}, {Wise}, {van Bemmel}, {Beck}, {Brienza}, {Bonafede}, {Calistro Rivera}, {Cassano}, {Clarke}, {Cseh}, {Deller}, {Drabent}, {van Driel}, {Engels}, {Falcke}, {Ferrari}, {Fr{\"o}hlich}, {Garrett}, {Harwood}, {Heesen}, {Hoeft}, {Horellou}, {Israel}, {Kapi{\'n}ska}, {Kunert-Bajraszewska}, {McKay}, {Mohan}, {Orr{\'u}}, {Pizzo}, {Prandoni}, {Schwarz}, {Shulevski}, {Sipior}, {Smith}, {Sridhar}, {Steinmetz}, {Stroe}, {Varenius}, {van der Werf}, {Zensus}, \& {Zwart}}]{2017shimwell}
{Shimwell}, T.~W., {R{\"o}ttgering}, H.~J.~A., {Best}, P.~N., {et~al.} 2017, \aap, 598, A104

\bibitem[{{Spergel} {et~al.}(2007){Spergel}, {Bean}, {Dor{\'e}}, {Nolta}, {Bennett}, {Dunkley}, {Hinshaw}, {Jarosik}, {Komatsu}, {Page}, {Peiris}, {Verde}, {Halpern}, {Hill}, {Kogut}, {Limon}, {Meyer}, {Odegard}, {Tucker}, {Weiland}, {Wollack}, \& {Wright}}]{2007spergel1}
{Spergel}, D.~N., {Bean}, R., {Dor{\'e}}, O., {et~al.} 2007, \apjs, 170, 377

\bibitem[{{Sulentic} \& {Marziani}(2015)}]{2015sulentic1}
{Sulentic}, J. \& {Marziani}, P. 2015, Frontiers in Astronomy and Space Sciences, 2, 6

\bibitem[{{Tortosa} {et~al.}(2022){Tortosa}, {Ricci}, {Tombesi}, {Ho}, {Du}, {Inayoshi}, {Wang}, {Shangguan}, \& {Li}}]{2022tortosa}
{Tortosa}, A., {Ricci}, C., {Tombesi}, F., {et~al.} 2022, \mnras, 509, 3599

\bibitem[{{Varglund} {et~al.}(2023){Varglund}, {J{\"a}rvel{\"a}}, {Ciroi}, {Berton}, {Congiu}, {L{\"a}hteenm{\"a}ki}, \& {Di Mille}}]{2023varglund}
{Varglund}, I., {J{\"a}rvel{\"a}}, E., {Ciroi}, S., {et~al.} 2023, \aap, 679, A32

\bibitem[{{Varglund} {et~al.}(2022){Varglund}, {J{\"a}rvel{\"a}}, {L{\"a}hteenm{\"a}ki}, {Berton}, {Ciroi}, \& {Congiu}}]{2022varglund}
{Varglund}, I., {J{\"a}rvel{\"a}}, E., {L{\"a}hteenm{\"a}ki}, A., {et~al.} 2022, \aap, 668, A91

\bibitem[{{Vietri} {et~al.}(2024){Vietri}, {Berton}, {J{\"a}rvel{\"a}}, {Kunert-Bajraszewska}, {Ciroi}, {Varglund}, {Dalla Barba}, {Sani}, \& {Crepaldi}}]{2024vietri}
{Vietri}, A., {Berton}, M., {J{\"a}rvel{\"a}}, E., {et~al.} 2024, \aap, 689, A123

\bibitem[{{Wang} {et~al.}(2016){Wang}, {Du}, {Hu}, {Bai}, {Wang}, {Yi}, {Wang}, {Zhang}, {Xin}, {Lun}, {Chang}, \& {Fan}}]{2016wang}
{Wang}, F., {Du}, P., {Hu}, C., {et~al.} 2016, \apj, 824, 149

\bibitem[{{Wang} {et~al.}(2023){Wang}, {Guo}, {Cao}, {Chen}, {Ding}, \& {Guo}}]{2023wang}
{Wang}, H., {Guo}, C., {Cao}, H., {et~al.} 2023, \apss, 368, 68

\bibitem[{{Wright}(2006)}]{2006wright}
{Wright}, E.~L. 2006, \pasp, 118, 1711

\bibitem[{{Xu} {et~al.}(2024){Xu}, {Komossa}, {Grupe}, {Wang}, {Xin}, {Han}, {Wei}, {Bai}, {Bon}, {Cangemi}, {Cordier}, {Dennefeld}, {Gallo}, {Kollatschny}, {Kong}, {Ochmann}, {Qiu}, \& {Schartel}}]{2024xu}
{Xu}, D.~W., {Komossa}, S., {Grupe}, D., {et~al.} 2024, Universe, 10, 61

\bibitem[{{Yang} {et~al.}(2020){Yang}, {Yao}, {Yang}, {Ho}, {An}, {Wang}, {Baan}, {Gu}, {Liu}, {Yang}, \& {Joshi}}]{2020yang}
{Yang}, X., {Yao}, S., {Yang}, J., {et~al.} 2020, \apj, 904, 200

\bibitem[{{Yuan} {et~al.}(2008){Yuan}, {Zhou}, {Komossa}, {Dong}, {Wang}, {Lu}, \& {Bai}}]{2008yuan}
{Yuan}, W., {Zhou}, H.~Y., {Komossa}, S., {et~al.} 2008, \apj, 685, 801

\bibitem[{{Zamanov} {et~al.}(2002){Zamanov}, {Marziani}, W., {Calvani}, {Dultzin-Hacyan}, \& {Bachev}}]{2002zamanov}
{Zamanov}, R., {Marziani}, P., W., S.~J., {et~al.} 2002, \apj, 576, 9

\bibitem[{{Zhou} {et~al.}(2006){Zhou}, {Wang}, {Yuan}, {Lu}, {Dong}, {Wang}, \& {Lu}}]{2006zhou}
{Zhou}, H., {Wang}, T., {Yuan}, W., {et~al.} 2006, \apj, 166, 128

\end{thebibliography}

\begin{appendix}

\section{Extreme}
\label{appextreme}

\begin{sidewaystable*}
\centering
\caption{Extreme behavior statistical limits}
\begin{tabular}{lllllllllllllllllllllll}
\hline\hline
Extendedness                                           & $\sigma$+ & $\sigma$- & N+ & N- &  &  &  &  &  &  &  &  &  &  &  &  &  &  &  &  &  &   \\
\hline\hline
VLASS Epoch 1                              &        & 2      &   & 16 &  &  &  &  &  &  &  &  &  &  &  &  &  &  &  &  &  &   \\
VLASS Epoch 2                              &        & 2      &   & 13 &  &  &  &  &  &  &  &  &  &  &  &  &  &  &  &  &  &   \\
FIRST                                      &        & 2      &   & 20 &  &  &  &  &  &  &  &  &  &  &  &  &  &  &  &  &  &   \\
                                                      &        &        &    &    &  &  &  &  &  &  &  &  &  &  &  &  &  &  &  &  &  &   \\
\hline
Radio luminosity                                      & $\sigma$+ & $\sigma$- & N+ & N- &  &  &  &  &  &  &  &  &  &  &  &  &  &  &  &  &  &   \\
\hline\hline
VLASS Epoch 1 Integrated flux density      & 1.5    & 1.5    & 14 & 9 &  &  &  &  &  &  &  &  &  &  &  &  &  &  &  &  &  &   \\
VLASS Epoch 1 Peak flux density            & 1.5    & 1.5    & 15 & 10 &  &  &  &  &  &  &  &  &  &  &  &  &  &  &  &  &  &   \\
VLASS Epoch 2 Integrated flux density      & 1.5    & 1.5    & 13 & 9 &  &  &  &  &  &  &  &  &  &  &  &  &  &  &  &  &  &   \\
VLASS Epoch 2 Peak flux density            & 1.5    & 1.5    & 14 & 9 &  &  &  &  &  &  &  &  &  &  &  &  &  &  &  &  &  &   \\
FIRST Integrated flux density              & 1.5    & 1.5    & 22 & 17 &  &  &  &  &  &  &  &  &  &  &  &  &  &  &  &  &  &   \\
FIRST Peak flux density                    & 1.5    & 1.5    & 21 & 16 &  &  &  &  &  &  &  &  &  &  &  &  &  &  &  &  &  &   \\
LoTSS Integrated flux density          & 2      & 2      & 45 & 30 &  &  &  &  &  &  &  &  &  &  &  &  &  &  &  &  &  &   \\
LoTSS Peak flux density                & 2      & 2      & 42 & 29 &  &  &  &  &  &  &  &  &  &  &  &  &  &  &  &  &  &   \\
                                         &        &        &    &    &  &  &  &  &  &  &  &  &  &  &  &  &  &  &  &  &  &   \\
\hline
Spectral index                 & $\sigma$+ & $\sigma$- & N+ & N- &  &  &  &  &  &  &  &  &  &  &  &  &  &  &  &  &  &   \\
\hline\hline
FIRST v. LoTSS peak           & 1.5      & 1.5      & 19  & 3 &  &  &  &  &  &  &  &  &  &  &  &  &  &  &  &  &  &   \\
FIRST v. LoTSS int            & 1.5      & 1.5      & 17  & 3  &  &  &  &  &  &  &  &  &  &  &  &  &  &  &  &  &  &   \\
VLASS Epoch 1 v. FIRST peak       & 1.5      & 1.5      & 11  & 4  &  &  &  &  &  &  &  &  &  &  &  &  &  &  &  &  &  &   \\
VLASS Epoch 1 v. FIRST int        & 1.5      & 1.5      & 13  & 3  &  &  &  &  &  &  &  &  &  &  &  &  &  &  &  &  &  &   \\
VLASS Epoch 2 v. FIRST peak       & 1.5      & 1.5      & 10  & 4  &  &  &  &  &  &  &  &  &  &  &  &  &  &  &  &  &  &   \\
VLASS Epoch 2 v. FIRST int        & 1.5      & 1.5      & 9   & 6  &  &  &  &  &  &  &  &  &  &  &  &  &  &  &  &  &  &   \\
                                &        &        &    &    &  &  &  &  &  &  &  &  &  &  &  &  &  &  &  &  &  &   \\
\hline
Variability                                           & $\sigma$+ & $\sigma$- & N+ & N- &  &  &  &  &  &  &  &  &  &  &  &  &  &  &  &  &  &   \\
\hline\hline
Fractional variability VLASS integrated flux density  & 2 &     & 8 &    &  &  &  &  &  &  &  &  &  &  &  &  &  &  &  &  &  &   \\
Fractional variability VLASS peak flux density        & 2 &     & 7 &    &  &  &  &  &  &  &  &  &  &  &  &  &  &  &  &  &  &   \\
Absolute change VLASS Epochs 1 and 2, integrated flux density & 1 & 1 & 5 & 5 &  &  &  &  &  &  &  &  &  &  &  &  &  &  &  &  &  &   \\
Absolute change VLASS Epochs 1 and 2, peak flux density       & 1 & 1 & 4  & 3  &  &  &  &  &  &  &  &  &  &  &  &  &  &  &  &  &  &   \\

\hline
\end{tabular}
\tablefoot{Columns: (1) Property type, (2) Positive statistical boundary, (3) Negative statistical boundary, (4) Number of sources above the positive boundary, (5) and Number of sources below the negative boundary.}
\label{tab:extreme}
\end{sidewaystable*}

\begin{sidewaysfigure}[htbp]
    \centering
    \adjustbox{valign=t}{\begin{minipage}{0.33\textwidth}
    \centering
    \includegraphics[width=1\textwidth]{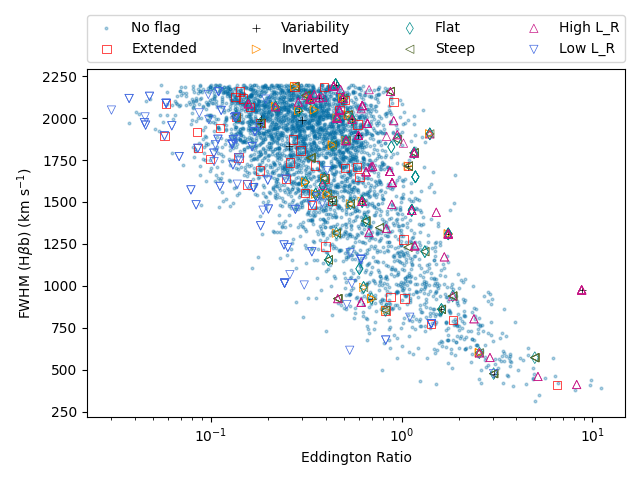}
    \end{minipage}}
    \adjustbox{valign=t}{\begin{minipage}{0.33\textwidth}
    \centering
    \includegraphics[width=1\textwidth]{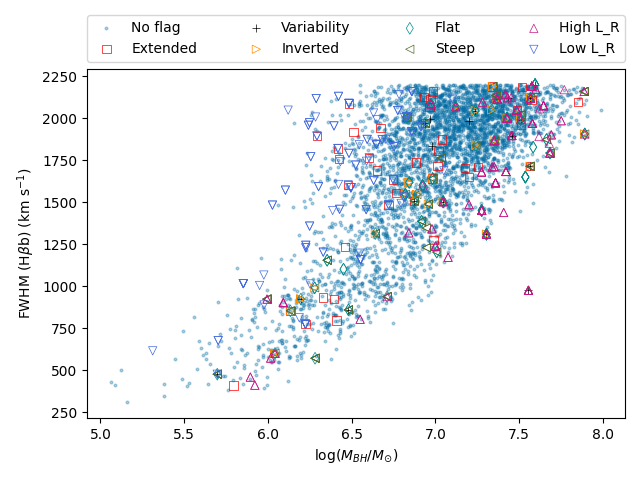}
    \end{minipage}}
    \adjustbox{valign=t}{\begin{minipage}{0.33\textwidth}
    \centering
    \includegraphics[width=1\textwidth]{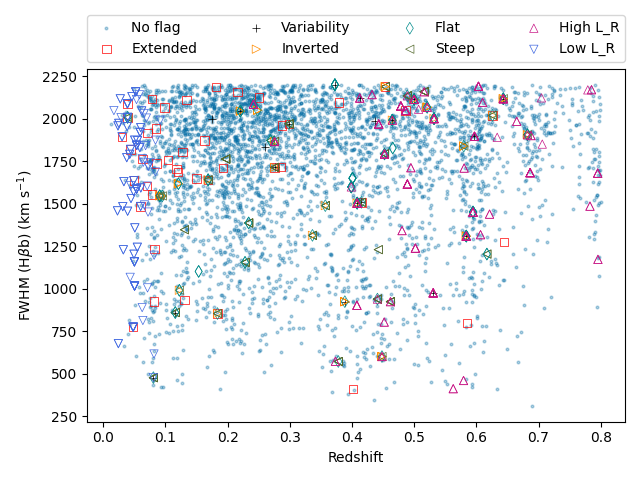}
    \end{minipage}}
    \hfill
        \caption{Extreme behavior of sources in FWHM(H$\beta$) plots. The blue, slightly opaque circles denote the sources with no extreme behavior. The red square symbols denote sources that are the most extended in our sample. The black plus symbols denote sources with extreme variability. The orange right-oriented triangles denote sources that have their highest spectral indices in the inverted region. The turquoise diamond symbols denote sources that have their highest spectral indices in the flat region. The dark green left-oriented triangles denote sources that have their lowest spectral indices in the steep region. The magenta right-side up triangles denote sources with the highest radio luminosities. The lilac upside-down triangles denote sources with the lowest radio luminosities. Spectral index extremes have been determined for 144~MHz - 1.4~GHz - 3~GHz (see Sect.~\ref{dataspecind}. \emph{Left panel:} Eddington ratio, \emph{middle panel:} Logarithmic black hole mass ratio compared with the solar mass, and \emph{right panel:} Redshift.}  \label{fwhm}
    \hfill
\end{sidewaysfigure}

\begin{sidewaysfigure}[htbp]
    \centering
\adjustbox{valign=t}{\begin{minipage}{0.33\textwidth}
\centering
\includegraphics[width=1\textwidth]{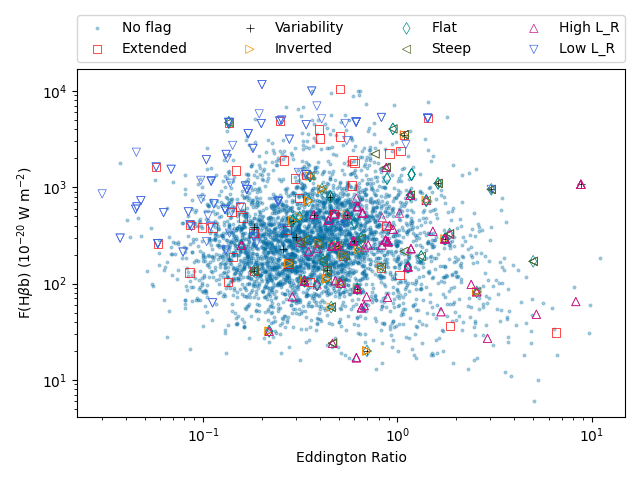}
\end{minipage}}
\adjustbox{valign=t}{\begin{minipage}{0.33\textwidth}
\centering
\includegraphics[width=1\textwidth]{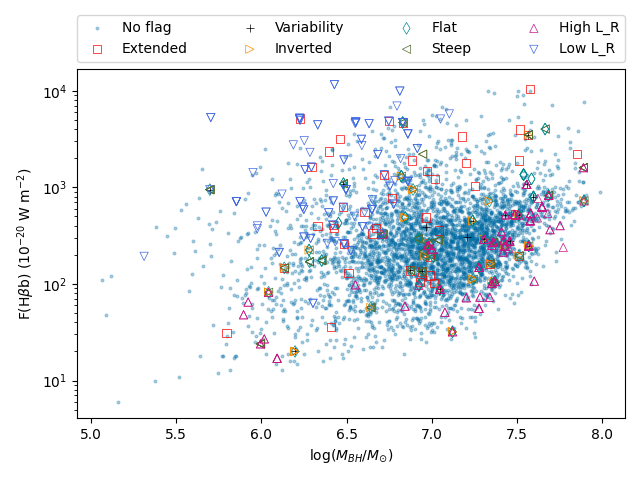}
\end{minipage}}
\adjustbox{valign=t}{\begin{minipage}{0.33\textwidth}
\centering
\includegraphics[width=1\textwidth]{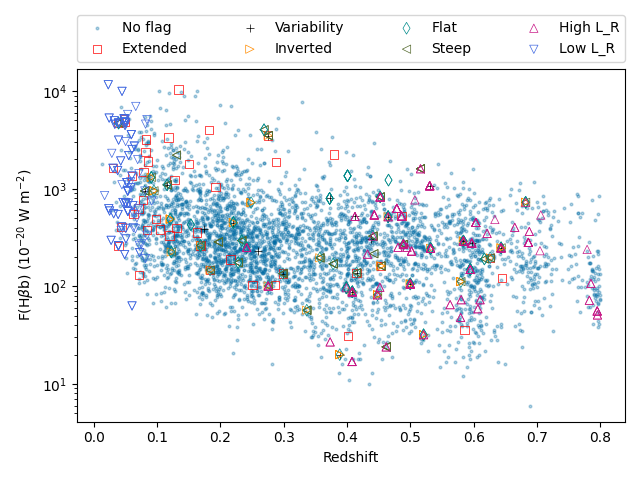}
\end{minipage}}
\hfill
    \caption{Extreme behavior of sources in FH($\beta$b) plots. Colors, symbols, and spectral indices as in Fig.\ref{fwhm}. \emph{Left panel:} Eddington ratio, \emph{middle panel:} Logarithmic black hole mass ratio compared with the solar mass, and \emph{right panel:} Redshift.}  \label{fhbb}
\hfill
\end{sidewaysfigure}

\begin{sidewaysfigure}
    \centering
\adjustbox{valign=t}{\begin{minipage}{0.33\textwidth}
\centering
\includegraphics[width=1\textwidth]{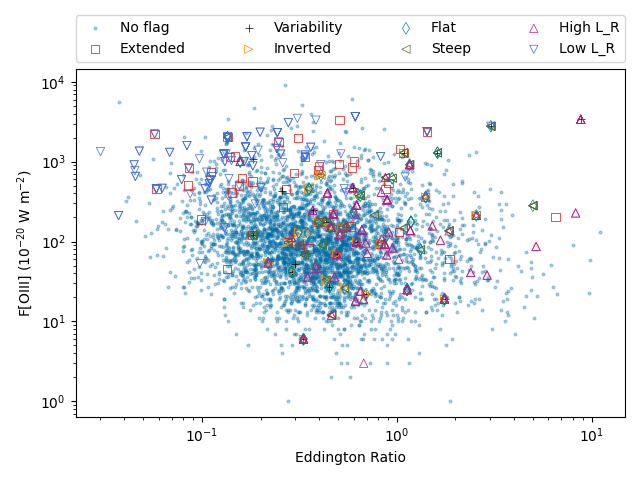}
\end{minipage}}
\adjustbox{valign=t}{\begin{minipage}{0.33\textwidth}
\centering
\includegraphics[width=1\textwidth]{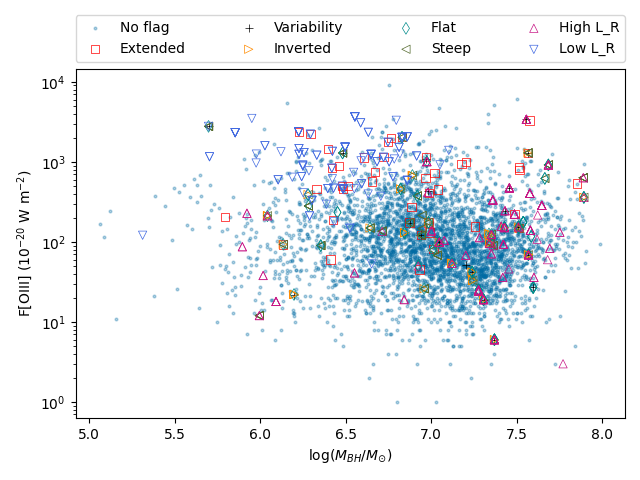}
\end{minipage}}
\adjustbox{valign=t}{\begin{minipage}{0.33\textwidth}
\centering
\includegraphics[width=1\textwidth]{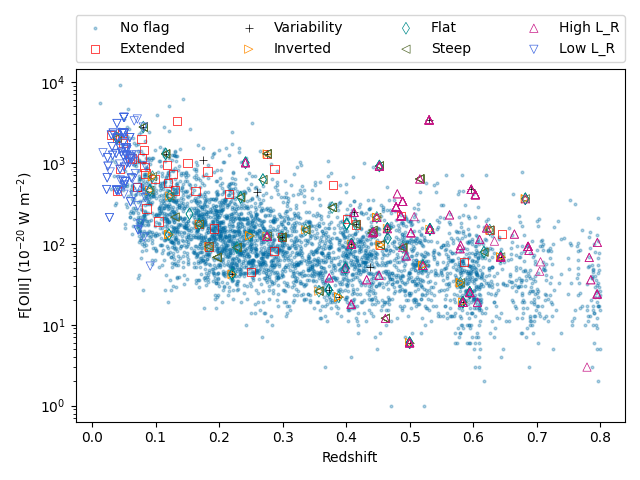}
\end{minipage}}
\hfill
    \caption{Extreme behavior of sources in F[O III] plots.  Colors, symbols, and spectral indices as in Fig.\ref{fwhm}. \emph{Left panel:} Eddington ratio, \emph{middle panel:} Logarithmic black hole mass ratio compared with the solar mass, and \emph{right panel:} Redshift.} \label{fo3}
\hfill
\end{sidewaysfigure}

\begin{sidewaysfigure}
    \centering
\adjustbox{valign=t}{\begin{minipage}{0.33\textwidth}
\centering
\includegraphics[width=1\textwidth]{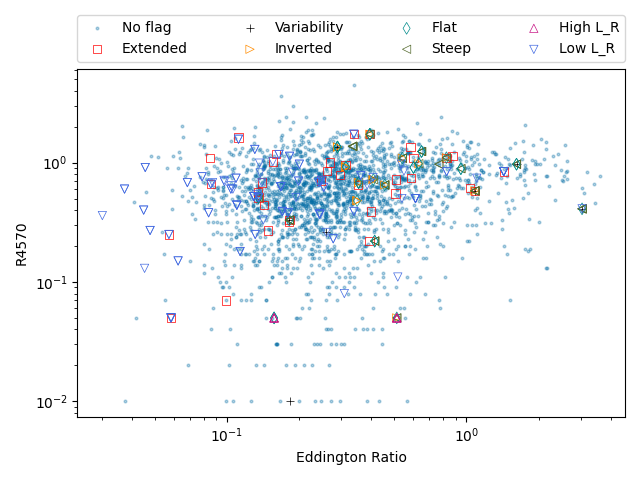}
\end{minipage}}
\adjustbox{valign=t}{\begin{minipage}{0.33\textwidth}
\centering
\includegraphics[width=1\textwidth]{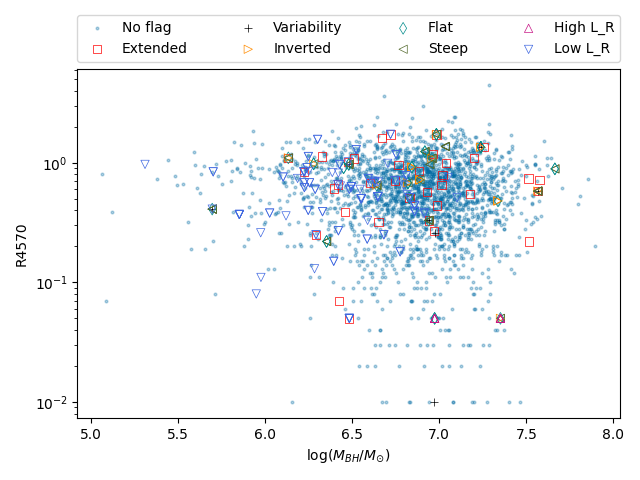}
\end{minipage}}
\adjustbox{valign=t}{\begin{minipage}{0.33\textwidth}
\centering
\includegraphics[width=1\textwidth]{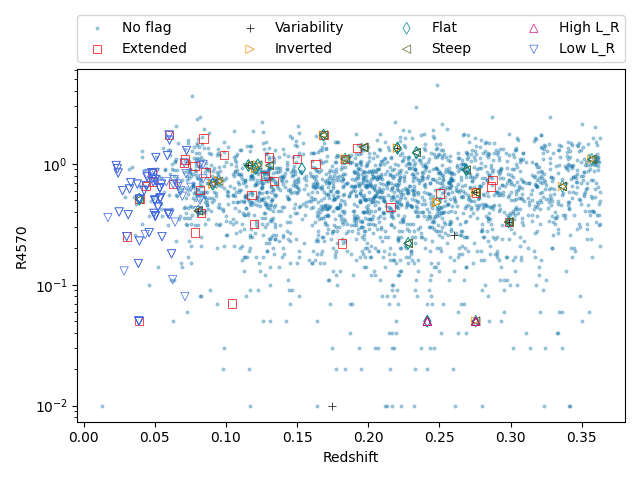}
\end{minipage}}
\hfill
    \caption{Extreme behavior of sources in R4570 plots. Colors, symbols, and spectral indices as in Fig.\ref{fwhm}. \emph{Left panel:} Eddington ratio, \emph{middle panel:} Logarithmic black hole mass ratio compared with the solar mass, and \emph{right panel:} Redshift.} \label{r4570}
\hfill
\end{sidewaysfigure}

\begin{sidewaysfigure}
    \centering
\adjustbox{valign=t}{\begin{minipage}{0.33\textwidth}
\centering
\includegraphics[width=1\textwidth]{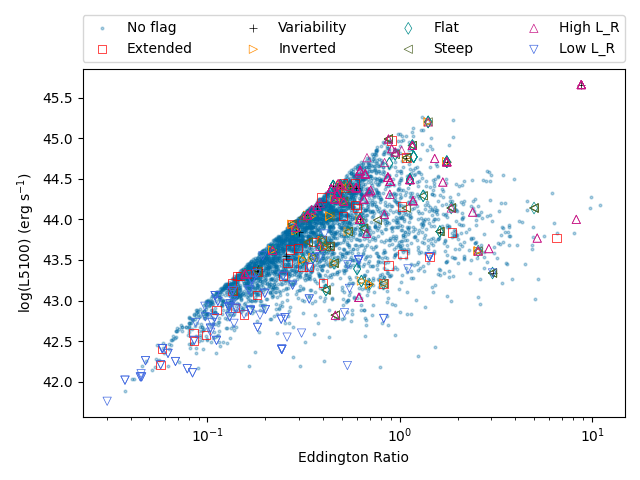}
\end{minipage}}
\adjustbox{valign=t}{\begin{minipage}{0.33\textwidth}
\centering
\includegraphics[width=1\textwidth]{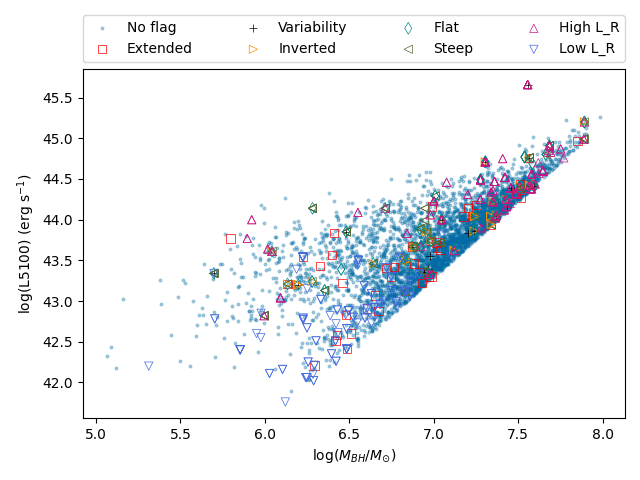}
\end{minipage}}
\adjustbox{valign=t}{\begin{minipage}{0.33\textwidth}
\centering
\includegraphics[width=1\textwidth]{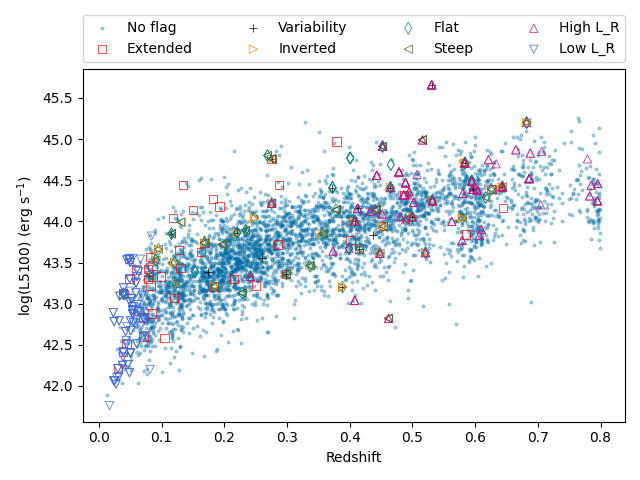}
\end{minipage}}
\hfill
    \caption{Extreme behavior of sources in L5100 plots. Colors, symbols, and spectral indices as in Fig.\ref{fwhm}. \emph{Left panel:} Eddington ratio, \emph{middle panel:} Logarithmic black hole mass ratio compared with the solar mass, and \emph{right panel:} Redshift.} \label{l5100}
\hfill
\end{sidewaysfigure}

\begin{sidewaysfigure}
    \centering
\adjustbox{valign=t}{\begin{minipage}{0.33\textwidth}
\centering
\includegraphics[width=1\textwidth]{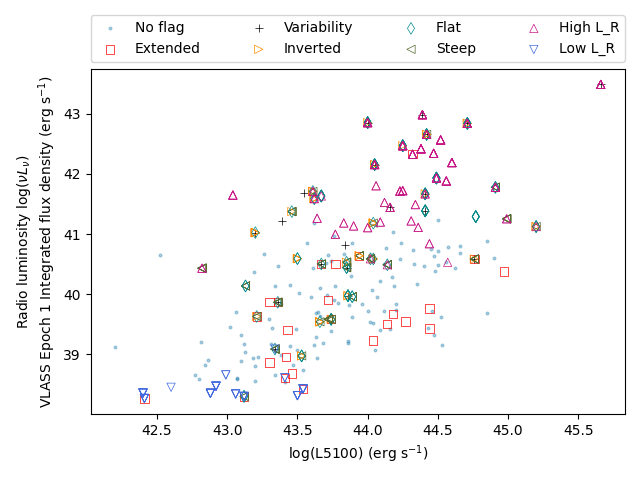}
\end{minipage}}
\adjustbox{valign=t}{\begin{minipage}{0.33\textwidth}
\centering
\includegraphics[width=1\textwidth]{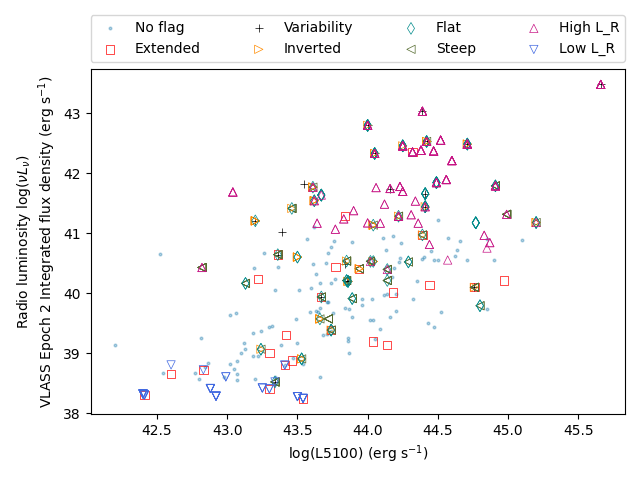}
\end{minipage}}
\hfill
   \caption{Radio luminosity and L5100 \emph{Left panel:} VLASS Epoch 1, and \emph{right panel:} VLASS Epoch 2. Colors, symbols, and spectral indices as in Fig.\ref{fwhm}.}  \label{rllum1l5100}
\hfill
\end{sidewaysfigure}

\begin{sidewaysfigure}
    \centering
\adjustbox{valign=t}{\begin{minipage}{0.33\textwidth}
\centering
\includegraphics[width=1\textwidth]{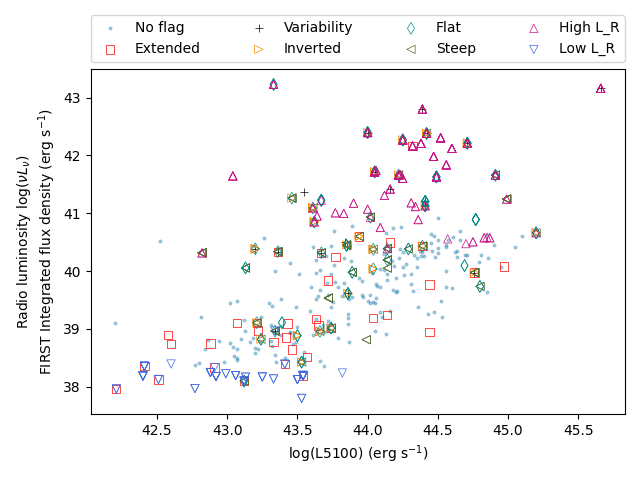}
\end{minipage}}
\adjustbox{valign=t}{\begin{minipage}{0.33\textwidth}
\centering
\includegraphics[width=1\textwidth]{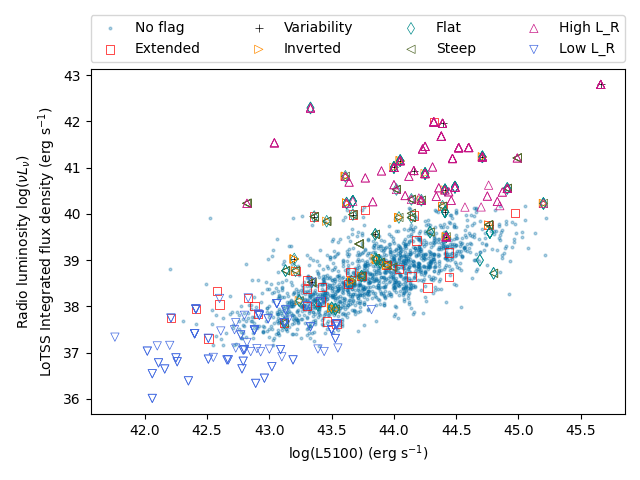}
\end{minipage}}
\hfill
    \caption{Radio luminosity and L5100 \emph{Left panel:} FIRST, and \emph{right panel:} LoTSS All. Colors, symbols, and spectral indices as in Fig.\ref{fwhm}.}  \label{rllum2l5100}
\hfill
\end{sidewaysfigure}

\begin{sidewaysfigure}
    \centering
\adjustbox{valign=t}{\begin{minipage}{0.33\textwidth}
\centering
\includegraphics[width=1\textwidth]{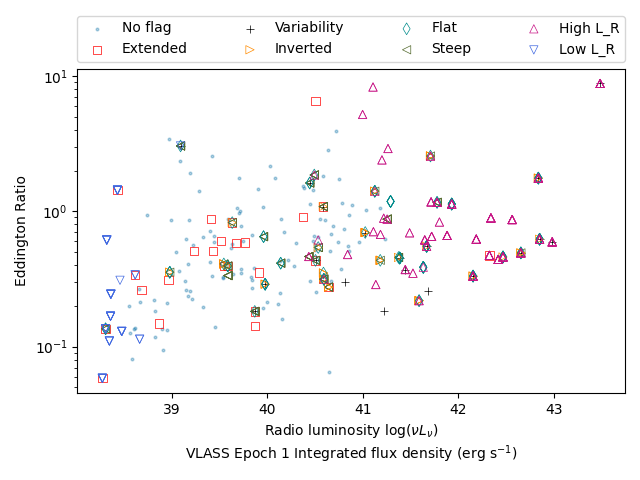}
\end{minipage}}
\adjustbox{valign=t}{\begin{minipage}{0.33\textwidth}
\centering
\includegraphics[width=1\textwidth]{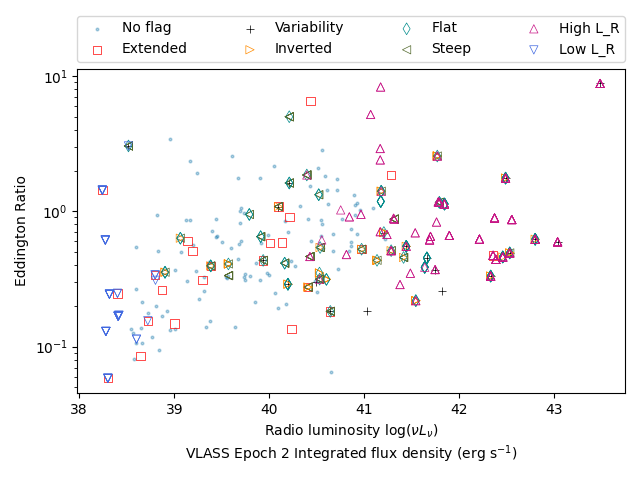}
\end{minipage}}
\hfill
    \caption{Radio luminosity and Eddington ratio \emph{Left panel:} VLASS Epoch 1, and \emph{right panel:} VLASS Epoch 2. Colors, symbols, and spectral indices as in Fig.\ref{fwhm}.}  \label{rllum1}
\hfill
\end{sidewaysfigure}

\begin{sidewaysfigure}
    \centering
\adjustbox{valign=t}{\begin{minipage}{0.33\textwidth}
\centering
\includegraphics[width=1\textwidth]{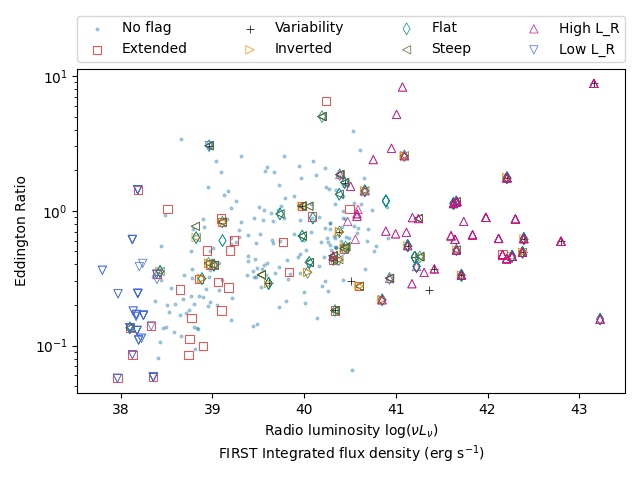}
\end{minipage}}
\adjustbox{valign=t}{\begin{minipage}{0.33\textwidth}
\centering
\includegraphics[width=1\textwidth]{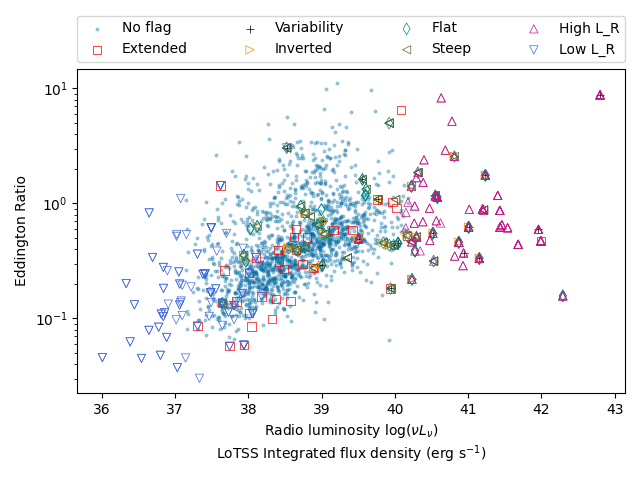}
\end{minipage}}
\hfill
    \caption{Radio luminosity and Eddington ratio \emph{Left panel:} FIRST, and \emph{right panel:} LoTSS All. Colors, symbols, and spectral indices as in Fig.\ref{fwhm}.}  \label{rllum2}
\hfill
\end{sidewaysfigure}

\begin{sidewaysfigure}
    \centering
\adjustbox{valign=t}{\begin{minipage}{0.33\textwidth}
\centering
\includegraphics[width=1\textwidth]{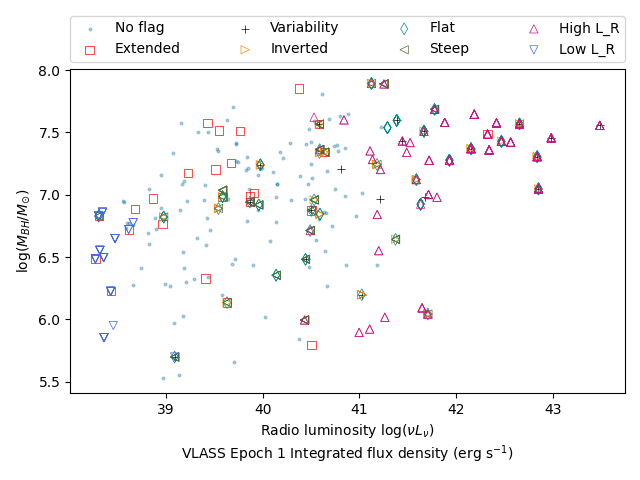}
\end{minipage}}
\adjustbox{valign=t}{\begin{minipage}{0.33\textwidth}
\centering
\includegraphics[width=1\textwidth]{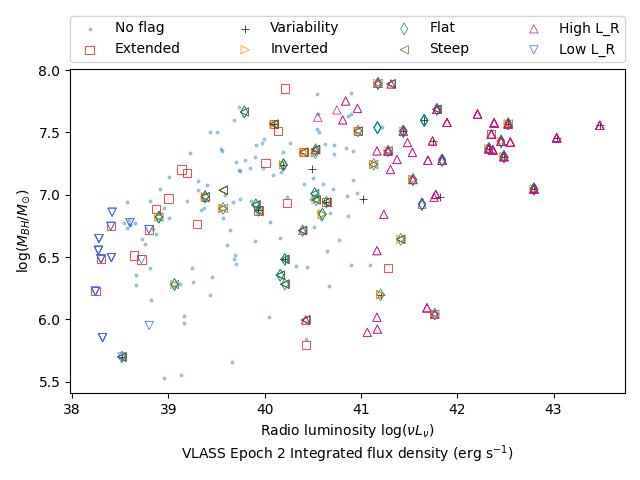}
\end{minipage}}
\hfill
    \caption{Radio luminosity and Mbh \emph{Left panel:} Epoch 1, and \emph{right panel:} Epoch 2. Colors, symbols, and spectral indices as in Fig.\ref{fwhm}.}  \label{rllumbh1}
\hfill
\end{sidewaysfigure}

\begin{sidewaysfigure}
    \centering
\adjustbox{valign=t}{\begin{minipage}{0.33\textwidth}
\centering
\includegraphics[width=1\textwidth]{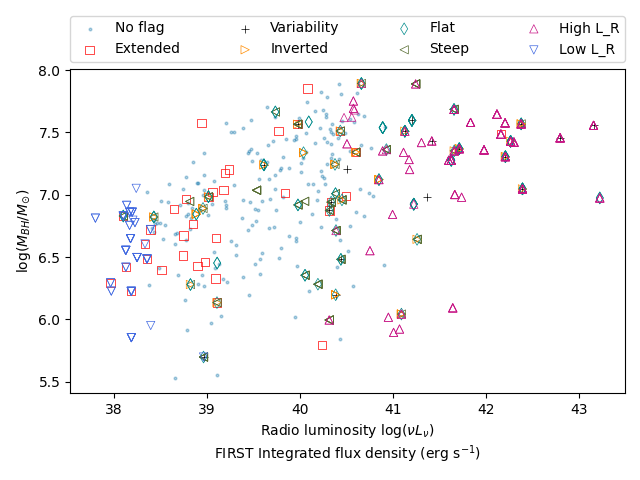}
\end{minipage}}
\adjustbox{valign=t}{\begin{minipage}{0.33\textwidth}
\centering
\includegraphics[width=1\textwidth]{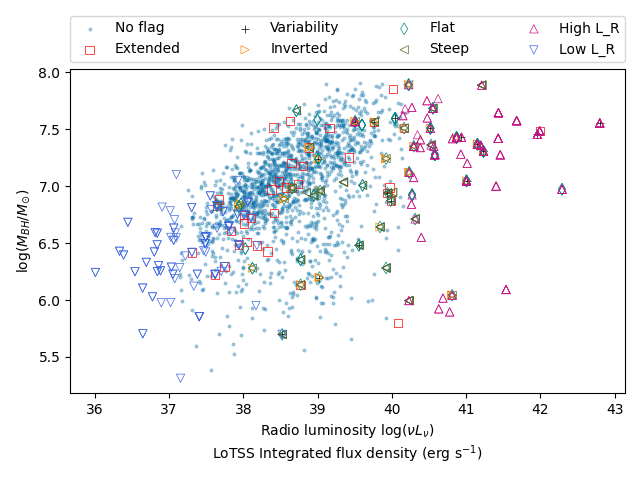}
\end{minipage}}
\hfill
    \caption{Radio luminosity and Mbh \emph{Left panel:} FIRST, and \emph{right panel:} LoTSS All. Colors, symbols, and spectral indices as in Fig.\ref{fwhm}.}  \label{rllummbh2}
\hfill
\end{sidewaysfigure}

\begin{sidewaysfigure}
    \centering
\adjustbox{valign=t}{\begin{minipage}{0.33\textwidth}
\centering
\includegraphics[width=1\textwidth]{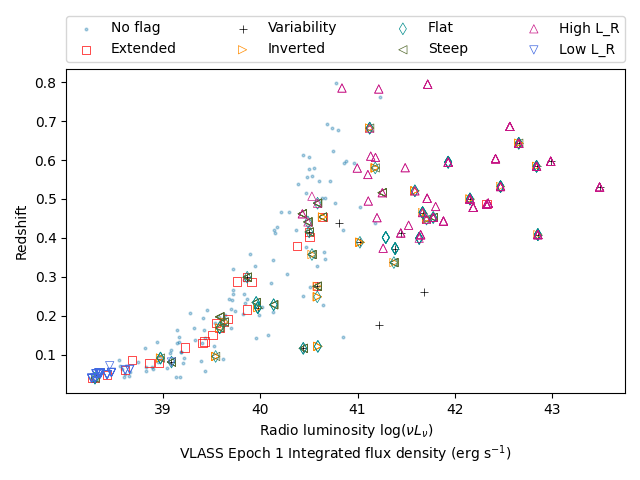}
\end{minipage}}
\adjustbox{valign=t}{\begin{minipage}{0.33\textwidth}
\centering
\includegraphics[width=1\textwidth]{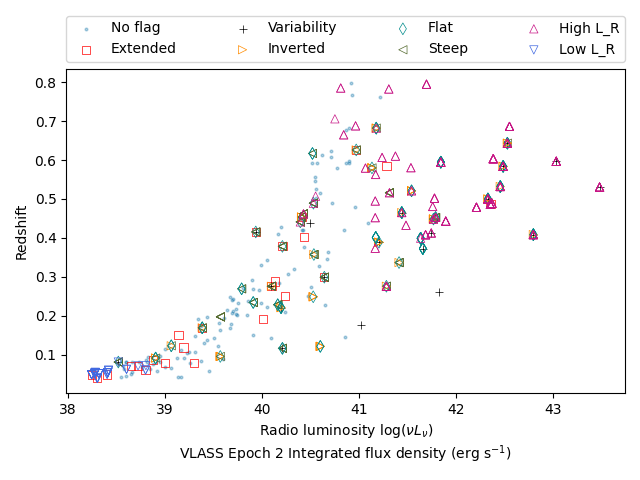}
\end{minipage}}
\hfill
    \caption{Radio luminosity and redshift \emph{Left panel:} Epoch 1, and \emph{right panel:} Epoch 2. Colors, symbols, and spectral indices as in Fig.\ref{fwhm}.}  \label{rllumz1}
\hfill
\end{sidewaysfigure}

\begin{sidewaysfigure}[htbp]
    \centering
\adjustbox{valign=t}{\begin{minipage}{0.33\textwidth}
\centering
\includegraphics[width=1\textwidth]{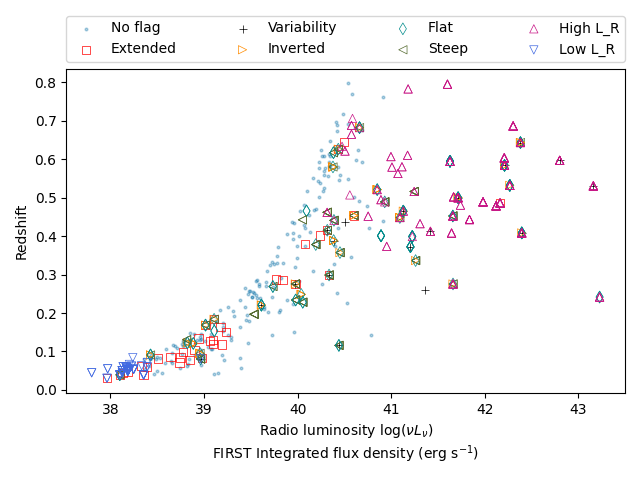}
\end{minipage}}
\adjustbox{valign=t}{\begin{minipage}{0.33\textwidth}
\centering
\includegraphics[width=1\textwidth]{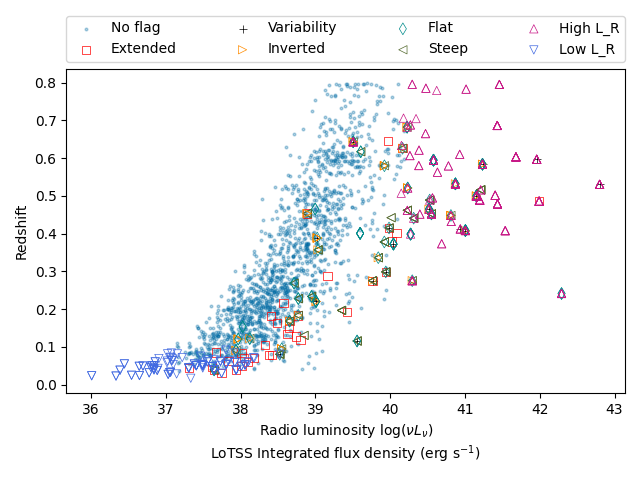}
\end{minipage}}
\hfill
    \caption{Radio luminosity and redshift \emph{Left panel:} FIRST, and \emph{right panel:} LoTSS All. Colors, symbols, and spectral indices as in Fig.\ref{fwhm}.}  \label{rllumz2}
\hfill
\end{sidewaysfigure}

\end{appendix}

\end{document}